\documentclass[12pt]{article}
\usepackage{natbib}
\usepackage{amsfonts,amssymb,amsmath,amsthm,bm,bigdelim,hyperref}
\usepackage{graphicx}
\usepackage{soul} 
\usepackage{multirow, multicol, float}
\usepackage{color}
\usepackage{rotating}
\usepackage[left=.8in,right=.8in,top=.8in,bottom=1in,footskip=0.2in]{geometry} 
\theoremstyle{plain}
\newtheorem{theorem}{Theorem}[section]

\newtheorem{proposition}[theorem]{Proposition}
\newtheorem{condition}[theorem]{Condition}

\usepackage{xr}
\begin{document}
\allowdisplaybreaks
\author{Changrui Liu and Solomon W.  Harrar\\Dr. Bing Zhang Department of Statistics\\
	University of Kentucky, Lexington, KY 40506, USA} 

\title{Wilcoxon-Mann-Whitney Effects for Clustered Data: Informative Cluster Size} 
\maketitle 
\abstract{
In clustered data setting, informative cluster size has been a focus of recent research.  In the nonparametric context, the problem has been considered mainly for testing equality of distribution functions.  The aim in this paper is to develop inferential procedure for the Wilcoxon-Mann-Whintey effect (also known as the nonprametric relative effect).  Unbiased estimator is provided and its asymptotic properties are investigated.  The asymptotic theory is employed to develop inferential methods.  While the proposed method takes information in the cluster sizes into consideration when constructing the estimator, it is equally applicable for ignorable cluster size situation.  Simulation results show that our method appropriately accounts for informative cluster size and it generally outperforms existing methods, especially those designed under ignorable cluster sizes.  The applications of the method is illustrated using data from a longitudinal study of alcohol use and a periodontal study.}



\section{Introduction}\label{sec:intro}

The Wilcoxon rank-sum test (also known as the Mann-Whitney $U$ test) \citep{W1945, MW1947} is a widely used nonparametric test for comparing two groups.  It is particularly suited for testing the hypothesis $H_0: p=1/2$, where 
\begin{equation}\label{eq:effectSize} 
	p=P(X_1 < X_2) +\frac{1}{2} P(X_1=X_2) 
\end{equation} 
is known as the Wilcoxon-Mann-Whitney (WMW) effect or the nonparametric relative effect,  and $X_1$ and $X_2$ are random variables representing observations from the two groups  \citep[see, e.g.,][]{FP1981,BM2000}.  The quantity $p$ is  interpreted as the tendency of observations made on $X_1$ to be smaller (larger) than those made on $X_2$ if $p >(<) 1/2$.  
The Wilcoxon rank-sum test assumes the sample observations to be mutually independent within and between groups.  Recently, several authors considered extensions to the clustered data setting \citep{RGL2004, DS2005, LHNO2010, DD2016, RHK2019, CKH2021}.  Some of these works allow members of the same cluster to be assigned in different groups for some of the clusters.

For testing the hypothesis
\begin{equation}
    H_0: p = \frac{1}{2}\label{hypstoc},
\end{equation}
 \citet{LHNO2010} proposed a weighting scheme and calculated a weighted average of the rank-sum statistic.  Subsequently, \citet{RHK2019} extended the approach with the options of using either weighted or unweighted estimators subject to specific research questions with enhanced interpretability.  However, their method is confined to the situation  where a  cluster cannot contain observations from both groups.  \citet{CKH2021} adapted a similar idea with more intuitive weighting scheme and allowed observations from both groups to be in the same cluster.  However, none of the above researches considered informative cluster size (ICS).  The goal of this article is to fill the gap in the literature by deriving a solution to testing problem \eqref{hypstoc} under ICS.

ICS is a condition where the size of a cluster may have an influence on the observations inside the cluster.  Under ICS, the distribution of an observation  depends on group membership as well as  the size of the cluster to which the observation belongs.  To give a concrete example, periodontal diseases  tend to be more prevalent at older ages.   Consider a study where interest lies whether a   factor (e.g. gum infections or smoking) affects the oral health of a subject over a period of time.  In this situation, observations taken from different teeth of the same  subject form a cluster.  When dealing with such data taking a tooth as a unit of analysis, we ought to take into account the fact that having fewer teeth would be an indication of poor oral health of a subject, assuming that there are no missing data; i.e.  cluster size is informative.  If each observation (tooth) is weighted equally, then subjects with fewer teeth would be down-weighted.  However, each subject should provide about the same amount of information.  Indeed, the number of remaining teeth provides important information about the overall oral health of a subject.  We will analyze a  periodontal data collected by \cite{BSKO1997} in Section \ref{sec:realdata}.

In cluster data setting, defining effect size requires some care.  Suppose $F_i(x)$ is the marginal distribution of a randomly selected subunit from a randomly selected cluster in group $i$.  The marginal analysis that is pursued in this paper estimates the WMW effect $p=\int F_1dF_2$.  Another marginal  analysis  used elsewhere, for example \citet{RHK2019} and  \citet{CKH2021}, targets $p_0$ which compare the marginal distribution of a randomly selected subunit from group 1 with that of a subunit from group 2 without regard to clusters. The interpretations of $p$ and $p_0$ are different in general, and they are the same only if cluster size is not informative.  For mathematical convenience, here and in the sequel we exclusively use the normalized distribution function which is  the average of the left and right continuous versions of the distribution function.

The remainder of this paper is organized as follows.  Section \ref{sec:motivation} gives a theoretical motivation for ICS.  In Section \ref{sec:estimation}, the within cluster resampling (WCR) method and the corresponding estimators for the WMW effect are presented.  Theoretical properties of the estimator are derived in Section \ref{sec:estimation}. Tests and confidence interval procedures are developed together with a small-sample approximation in Section \ref{sec:test}.  Simulation results showing the finite sample performance of the proposed estimation and test procedures in comparison with popular and recent methods in the literature is the subject of Section \ref{sec:sim}.  In Section \ref{sec:realdata}, detailed analyses of two real datasets are presented.  Further discussions and conclusions are provided in Section \ref{sec:conclusion}.  All proofs and additional simulation results are contained in the Supplementary Material.



\section{Informative Cluster Size}\label{sec:motivation}

To illustrate the issues with informative (nonignorable) cluster size, let cluster size have  uniform distribution over $\{c_1,c_2\}$, where $c_1$ and $c_2$ are positive integers.  Let   $F(x|g=0,m=c_1)=\Phi(x+c_2)$, $F(x|g=0,m=c_2)=\Phi(x-c_1)$, $F(x|g=1,m=c_1)=\Phi(x-c_2)$, and $F(x|g=1,m=c_2)=\Phi(x+c_1)$, where  $F(x|g, m)$ is the  (conditional) distribution of an observation in group $g$ and from a cluster with size $m$ and $\Phi(\cdot)$ denotes the CDF of the standard normal distribution.  For the sake of simplicity, we assume that there are no complete clusters.

\begin{figure}[H]
	\centering
	\includegraphics[scale = 0.7]{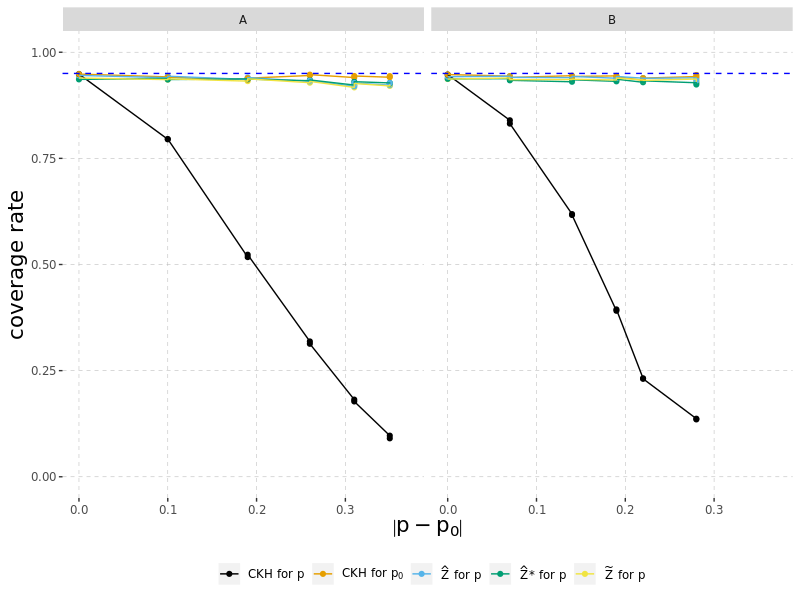}
	\caption{\label{fig:introcomparison} Coverage rates of 95\% confidence intervals  plotted against the absolute difference between $p$ and $p_0$, which are effect sizes under informative and ignorable, respectively, cluster sizes. CKH is the method in \citet{CKH2021} and  $\widetilde Z$ is the method proposed in this paper.  The methods $\widehat Z$ and $\widehat Z^*$ are discussed in Section \ref{sec:test}.  Panel A is for Gaussian distribution and  panel B is for Cauchy distribution.  Dotted blue line shows 0.95.}
\end{figure}

One can easily see that 
$$F_g(x) = \frac12\Phi(x-c_g) + \frac12\Phi(x+c_{g'})$$ for $g,g'\in \{1,2\}$ and $g\ne g'$.  Note that the  distribution function $F_1$ and $F_2$ are different if $c_1\ne c_2$, and the  treatment effect $p$ is
\begin{align*}
	p &= \int F_1dF_2= \frac{1}{2}\Phi\bigg(\frac{c_2-c_1}{\sqrt{2}}\bigg) + \frac{1}{4}\Phi(\sqrt{2}c_2) + \frac{1}{4}\Phi(-\sqrt{2}c_1).
\end{align*}
Disregarding the information in the cluster sizes, the distribution of an observation in group $g$, denoted by $F^{(0)}_g$, is
$$F_g^{(0)}(x) = \frac{c_{g'}}{c_g+c_{g'}}\Phi(x-c_g) + \frac{c_{g}}{c_g+c_{g'}}\Phi(x+c_{g'}),$$
for $g,g'\in \{1,2\}$ and $g\ne g'$.  The parameter of  interest in this model, denoted by $p_0$, is  
\begin{align*}
	p_0 &= \int F_1^{(0)}dF_2^{(0)}= \frac{2c_1c_2}{(c_1+c_2)^2}\Phi\left(\frac{c_2-c_1}{\sqrt{2}}\right) + \left(\frac{c_1}{c_1+c_2}\right)^2 \Phi(\frac{c_2}{\sqrt{2}}) + \left(\frac{c_2}{c_1+c_2}\right)^2\Phi(-\frac{c_1}{\sqrt{2}}).
\end{align*}
Table \ref{table1} shows that $p$ and $p_0$ take increasingly different values as the difference between $c_1$ and $c_2$ increases.  The two probabilities are equal if $c_1=c_2$, in which case the cluster size is not informative.  It should be emphasized that methods that assume ignorable cluster size \citep[e.g.,][]{LHNO2010, RHK2019, CKH2021} are aiming at $p_0$.
\begin{table}[H]
	\begin{center}
		\begin{tabular}{||c|c|c|c|c|c|c|c|c|c|c|c||} 
			\hline
			$c_1$ & 2 & 2 & 3 & 2 & 4 & 2 & 5 & 2 & 6 & 2 & 7 \\[0.2ex]
			\hline
			$c_2$ & 2 & 3 & 2 & 4 & 2 & 5 & 2 & 6 & 2 & 7 & 2 \\
			\hline
			$p$ & 0.50 & 0.63 & 0.37 & 0.71 & 0.29 & 0.74 & 0.26 & 0.75 & 0.25 & 0.75 & 0.25\\
			\hline
			$p_0$ & 0.50 & 0.53 & 0.47 & 0.52 & 0.48 & 0.48 & 0.52 & 0.44 & 0.56 & 0.40 & 0.60\\
			\hline
		\end{tabular}\\
		\caption{\label{table1} Wilcoxon-Mann-Whitney  effect under informative ($p$) and ignorable ($p_0$) cluster size.  The parameters $c_1$ and $c_2$ control the informativeness of cluster size. The case $c_1=c_2$ is noninformative (ignorable) cluster size.} 
	\end{center}
\end{table}
Figure \ref{fig:introcomparison} plots coverage probabilities of confidence intervals for $p$ as a function of $|p-p_0|$.  We are showing the methods by \cite{CKH2021} {(abbreviated as ``CKH" hereinafter)} and the methods based on  ${\widehat p}^*$, $\widehat p$ and $\widetilde p$ proposed later in this paper.  When $c_1=c_2$ , i.e. $p_0=p$,   the quantities estimated by the different methods  coincide.  However, in the other cases, the parameter estimates are different and the new methods are appropriate.  While CKH performs poorly for estimating $p$, its coverage for $p_0$ is quite well.


\section{Estimation}\label{sec:estimation}

Let  $X_{ij}$ denote the outcome variable measured from the $j^{th}$ subunit in the the $i^{th}$ cluster, $i=1,\cdots,n$ and $j=1,\cdots,m_i$, where $m_i$ denote the size of the $i^{th}$ cluster.  Further, let $g_{ij}$ denote the binary group membership indicator of the $j^{th}$ observation from the $i^{th}$ cluster, i.e., $g_{ij} = 0$ if $X_{ij}$ is in group 1 and $g_{ij}=1$ if $X_{ij}$ is in group 2.  Let $\bm{X}_i=(X_{i1},\dots,X_{im^{(1)}},X_{im^{(1)}+1},\ldots,X_{im_i})^\top$, $\bm{g}_i=(g_{i1},\dots,g_{im^{(1)}},g_{im^{(1)}+1},\ldots,g_{im_i})^\top$ and $\bm{X}=(\bm{X}_1^\top,\ldots,\bm X_n^\top)^\top$ and $\bm{g}=(\bm{g}_1^\top,\ldots,\bm g_n^\top)^\top$.  The notation $(\bm{X},\bm{g}) = \{(X_{ij},g_{ij}): i=1,2,\cdots,n;$ $j = 1,2,\cdots,m_i\}$ will represent the  full data from the two groups.  We make the assumption that $g_{ij}$ is fixed for all $i=1,\cdots,n$ and $j=1,\cdots,m_i$.  Let $n_1$ denote the number of clusters that contain only observations from group 1 and none from group 2 (incomplete clusters in group 1), $n_2$ denote the number of clusters that contain only observations from group 2 and none from group 1 (incomplete clusters in group 2), and $n_c$ denote the number of clusters that contain observations from both groups (complete clusters).  Note that $n = n_1+n_2+n_c$.  For convenience, assume that the data is arranged so that $i=1,\ldots, n_1$ corresponds to the incomplete clusters in groups 1, $i=n_1+1,\ldots, n_1+n_2$ corresponds to the incomplete clusters in group 2, and $i=n_1+n_2+1,\ldots, n$ corresponds to the complete clusters.  We further define $R_1=\{1,\ldots,n_1,n_1+n_2+1,\ldots,n\}$ and $R_2=\{n_1+1,\ldots,n\}$ to be sets of indices for clusters in groups 1 and group 2, respectively. Notice that  $m_i^{(2)}=0$ for $i=1,\ldots,n_1$ and $m_i^{(1)}=0$ for $i=n_1+1\ldots,n_1+n_2$.
Inside each complete cluster $i$, we assume that observations are arranged such that 
$X_{i1},\cdots,X_{im_i^{(1)}}$ are from group 1 and $X_{i,m_i^{(1)}+1},\cdots,X_{im_i}$ are from group 2, where $m^{(1)}_{i} = m_i - m^{(2)}_{i}$ and $m^{(2)}_{i} = \sum_{j=1}^{m_i} g_{ij}$ denote the number of observations belonging to group 1 and 2, respectively, within cluster $i$ (intra-cluster group sizes).  This arrangement guarantees that $g_{ik} = 0$ for $k=1,\cdots,m_i^{(1)}$ and $g_{ik} = 1$ for $k=m_i^{(1)}+1,\cdots,m_i$.    

\subsection{Ignorable Cluster Size} \label{subsec:ICS}
To motivate our estimators, we briefly discuss estimation methods when  cluster size is not informative. Let $G_\ell^{(\phi)}$ for $\phi \in \{u,w\}$ be the  unweighted and weighted, respectively, empirical distribution function of observations in group $\ell$, i.e., 
\begin{align*}
	G_1^{(u)}(x) &= \frac{1}{n_1+n_c} \sum_{i\in R_1}\frac{1}{m_i^{(1)}}\sum_{j=1}^{m_i}(1-g_{ij})\bigg[\frac{1}{2}\bigg(I(X_{ij}< x) + I(X_{ij}\le x)\bigg)\bigg],\\
	G_2^{(u)}(x) &= \frac{1}{n_2+n_c} \sum_{i \in R_2}\frac{1}{m_i^{(2)}}\sum_{j=1}^{m_i}g_{ij}\bigg[\frac{1}{2}\bigg(I(X_{ij}< x) + I(X_{ij}\le x)\bigg)\bigg],\\
	G_1^{(w)}(x) &= \frac{1}{N_1} \sum_{i\in R_1}\sum_{j=1}^{m_i}(1-g_{ij})\bigg[\frac{1}{2}\bigg(I(X_{ij}< x) + I(X_{ij}\le x)\bigg)\bigg]\quad \text{and}\\
	G_2^{(w)}(x) &= \frac{1}{N_2} \sum_{i\in R_2}\sum_{j=1}^{m_i}g_{ij}\bigg[\frac{1}{2}\bigg(I(X_{ij}< x) + I(X_{ij}\le x)\bigg)\bigg],
\end{align*} 
where $N_1 = \sum_{i=1}^{n} m_i^{(1)}$ and $N_2 = \sum_{i=1}^{n} m_i^{(2)}$ denote the total number of observations from group 1 and 2, respectively, and $I(\cdot)$ is the indicator function.

When cluster size is noninformative or ignorable, one may use the estimators $\widehat p^{(\phi)}=\int G_1^{(\phi)}dG_2^{(\phi)}$ for $\phi \in \{u,w\}$ \citep{RHK2019, CKH2021} or other weighted estimators \citep{LHNO2010} for $p$.   In these papers, unbiasedness, consistency and asymptotic normality of the estimators are thoroughly investigated.  However, as discussed in Section \ref{sec:motivation}, the quantities estimated by these estimators have quite different interpretations and are not appropriate under informative cluster size.  { In the sequel, $G_1$ and $G_2$ without superscript refer to $G_1^{(u)}$ and $G_2^{(u)}$, respectively, unless otherwise specified.}

\subsection{Proposed Estimator}\label{subsec:estimators}

To overcome the difficulty with informative cluster size discussed in Section \ref{sec:motivation}, we propose a within cluster resampling (WCR) based estimator.  The WCR was first proposed by \cite{HSW2001} in a GEE context \citep[see also][]{WDS2003} and  later used for the rank-sum test  by \citet{DS2005} and others \citep[e.g.,][]{DS2008, DD2016}. In WCR estimation, one random observation is drawn from  each cluster and an estimator of the quantity of interest is constructed based on this resample.  The  average over all possible resamples are taken to get the final estimator.

For a given resample $\bm{X^*} = (X^*_{1}, X^*_{2}, \cdots, X^*_{n})$ and $\bm{g}^* = (g^*_1, \cdots, g^*_n)$, let $m_1^* = \sum_{i=1}^{n} (1-g^*_{i})$ and $m_2^* = n - m_1^*$ denote the number of observations coming from group 1 and 2, respectively.  In addition, denote the normalized version of empirical distribution functions of observations in group 1 and 2 as $\widehat{F}_1^*$ and $\widehat{F}_2^*$, respectively.  
An estimator $\widehat{p}^*$ of $p$ based on the resample $(\bm{X^*,g^*})$ is 
    \begin{align*}
        \widehat{p}^* &= \int \widehat{F}_1^* d\widehat{F}_2^*
        = \frac{1}{m_1^*m_2^*} \sum_{i=1}^{n}\sum_{j\ne i} (1-g_i^*)g_j^*\cdot\frac{1}{2}\bigg[I(X^*_{i} < X^*_{j}) +  I(X^*_{i} \le X^*_{j})\bigg]
        = \frac{U^*}{m_1^*m_2^*},
    \end{align*}
where
\begin{align*}
    U^*&= \sum_{i=1}^{n} g_i^*R_i^* - \frac{m_2^*(m_2^*+1)}{2}\quad \text{and} \quad 
R_i^* = 1 + \frac{1}{2} \bigg[\sum_{j\ne i}I(X_j^* < X_i^*) + \sum_{j\ne i}I(X_j^*\le X_i^*)\bigg].\end{align*}
This is the same estimator as \citet{BM2000} {treating the resample as a dataset from two independent samples}.  Note that  $U^*$ is the Mann-Whitney U-statistic and $R_i^*$ is 
is the mid rank of $X_i^*$ within the resample $\boldsymbol{X}^* = (X_1^*, \cdots, X_n^*)$.   
 The estimator $\widehat{p}^*$ itself is inefficient because it is based on only a small part of the full data.  A more reasonable estimator would be 
\begin{align}\label{eq:p_hat}
    \widehat{p} = E(\widehat{p}^*|\bm{X}, \bm{g}),
\end{align}
where the estimates $\widehat p^*$ are averaged over all possible resamples.  However,  $\widehat{p}$ does not have a closed-form expression.  It needs to  be  evaluated numerically by averaging $\widehat{p}^*$ over a large number, say $Q = 10,000$, of resamples.  Monte Carlo process, besides its computational burden,  adds a layer of uncertainty which may affect the precision of the estimator.  To overcome these problems, we   consider averaging the rank sum statistic $U^*$ over the resampling distribution.  It is not difficult to see \citep[e.g.,][]{DS2005}  that 
\[E\left(U^*|\bm{X},\bm{g}\right)= n\sum_{i=1}^{n}\alpha_i - \sum_{i=1}^{n}\sum_{j\ne i}\sum_{k=1}^{m_j} \frac{\alpha_i}{m_j} \widehat{F}_{2i}(X_{jk}) - \frac{1}{2}\left[\sum_{i=1}^{n}\alpha_i(1-\alpha_i)+ \left(\sum_{i=1}^{n}\alpha_i\right)^2+\sum_{i=1}^{n}\alpha_i\right],\] 
where  $\widehat{F}_{2i}$ denotes the normalized version of the empirical distribution function of observations in group 2 within cluster $i$ and $\alpha_i = m_i^{(2)}/m_i$ denotes the proportion of observations coming from group 2 within cluster $i$.   For a given resample $\bm X^*$, the statistic $U^*$  roughly counts the number of times a member from group 1 is smaller than a member from group 2.  To normalize $E(U^*|\bm X,\bm g)$, we consider  the average number of possible comparisons \[E(m_1^*m_2^*|\bm X,\bm g)= \sum_{i=1}^n\sum\limits_{j\ne i} \alpha_i(1-\alpha_j).\] Therefore, we propose to  estimate  $p$  by 
\begin{equation}\label{eq:p_tilde}
    \widetilde{p} = \frac{E(U^*|\bm{X,g})}{E(m_1^*m_2^*|\bm{X}, \bm{g})} = \frac{1}{2} + \frac{\frac{n-1}{2}\sum\limits_{i=1}^{n} \alpha_i - \sum\limits_{i=1}^{n}\sum\limits_{j\ne i}\sum\limits_{k=1}^{m_j} \frac{\alpha_i}{m_j} \widehat{F}_{2i}(X_{jk})}{\sum\limits_{i=1}^{n}\sum\limits_{j\ne i}\alpha_i(1-\alpha_j)}.
\end{equation}
It is worth pointing out that in the case when there are no complete clusters, i.e. $n_c=0$, $m_1^*=n_1$ and $m_2^*=n_2$ and, thus, the estimators $\widetilde p$ and $\widehat p$ coincide. 

\subsection{Properties of the Estimator}\label{subsec:properties}

The asymptotic results presented in this paper need the following technical conditions.
 \begin{condition}\label{cond:numclus} The sample sizes (number of clusters) $n_c$, $n_1$ and $n_2$ satisfy
 	$$\min(n_1+n_c, n_2+n_c)\rightarrow \infty \text{ as } n\rightarrow\infty.$$
 \end{condition}

\begin{condition}\label{cond:regvar}
	The distribution functions $F_1$ and $F_2$ are non-degenerate.
	\end{condition}

\begin{condition}\label{cond:finclussize}
	There exists some positive integer $K$ such that $\sup\limits_{1\le i\le n}m_i \le K$ for any $n$.
\end{condition}

Condition \ref{cond:numclus} requires that the number of clusters in each group is large.  It is satisfied if either one of the following two scenarios is satisfied:
\begin{itemize}
	\item[(i)] $n_c\rightarrow\infty$ as $n\rightarrow\infty$, or
	\item[(ii)] $n_1\rightarrow\infty$ and $n_2\rightarrow\infty$ as $n\rightarrow\infty$.
\end{itemize}
Condition \ref{cond:regvar} guarantees that 
\begin{align*}
	\inf_{i,j} Var[F_i(X_{ij})] >0 \text{ and } \inf_{i,j} Var[F_2(X_{ij})] >0,
\end{align*} for any $n$. This condition is   needed for the Lindeberg's Condition of the Central Limit Theorem.  Condition \ref{cond:finclussize}  is rather mild in the sense that the number of subunits per cluster is bounded which is typically the case in practical applications.

\begin{proposition}\label{prop:ptilde}
	\begin{itemize} The estimator $\widetilde p$ is unbiased and, under Condition \ref{cond:numclus}, consistent.  Specifically, 
		\item[(i)]  $E(\widetilde{p}) = p$ and
		\item[(ii)] under Condition \ref{cond:numclus}, $\widetilde{p}\stackrel{p}{\to} p$ as $n\rightarrow\infty$.
	\end{itemize}
\end{proposition}



\section{Hypothesis Test and Confidence Interval}\label{sec:test}
\subsection{Asymptotic Procedures}\label{subsec:testintro}
Asymptotic test and confidence interval can be constructed in the usual way if we have the asymptotic distribution of $\widetilde p$.
\begin{theorem}[Asymptotic Normality]\label{thm:ptildenorm}
	Under {Conditions \ref{cond:numclus}, \ref{cond:regvar} and \ref{cond:finclussize}},
	\begin{align*}
		\frac{\widetilde{p} - p}{\sqrt{{\rm Var}(\widetilde{p})}} \stackrel{d}{\rightarrow} N(0,1)
	\end{align*}
	as $n\rightarrow\infty$,  where     
        \begin{align*}
		{\rm Var}(\widetilde{p}) = \left(\frac{n+1}{\sum_{i=1}^{n}\sum_{j\ne i}\alpha_i(1-\alpha_j)}\right)^2\sum_{i=1}^{n}{\rm Var}(W_i),
	\end{align*}
\begin{align*}
	W_i &= -\frac{1}{nm_i}\sum_{j=n_1+1,j\ne i}^{n}\alpha_j\sum_{k=1}^{m_i}F_2(X_{ik}) +\frac{{\rm I}(i\in R_2)}{nm_i}\sum_{j=1,j\ne i}^{n}\frac{1}{m_j}\sum_{k=1}^{m_i}\sum_{h=1}^{m_j}g_{ik}f_{jh}(X_{ik}),
\end{align*}
and 
$f_{jh}(x) = E\left[\frac{1}{2}\left(I(X_{jh}<X_{ik})+I(X_{jh}\le X_{ik})\right)|X_{ik}=x\right]$.
\end{theorem}

\begin{theorem}[Ratio Consistency]\label{thm:ratioconsistency}
Under Conditions \ref{cond:numclus}, \ref{cond:regvar} and \ref{cond:finclussize},
$$\frac{\widehat{{\rm Var}(\widetilde{p})}}{{\rm Var}(\widetilde{p})} \stackrel{p}{\to} 1$$ as $n\to \infty$, where
\begin{align*}
	\widehat{{\rm Var}(\widetilde{p})} =  \left(\frac{n+1}{\sum_{i=1}^{n}\sum_{j\ne i}\alpha_i(1-\alpha_j)}\right)^2\sum_{l=1}^{n}(\widehat{W}_l - \widehat{E(W_l)})^2,
\end{align*}
\begin{align*}
	\widehat{W}_l = \frac{1}{m_l(n+1)}\left(\sum_{i=n_1+1,i\ne l}^{n}\alpha_i\sum_{k=1}^{m_l} G_2(X_{lk}) -{\rm I}(l \in R_2) \sum_{j=1, j\ne l}^{n}\sum_{h=m_l^{(1)}+1}^{m_l} \widehat{F}_j(X_{lh})\right)
\end{align*}
and
\begin{align*}
\widehat{E(W_l)}&= \frac{1}{n+1}\left[\left((1-\alpha_l)(1-\widetilde{p}) + \frac{\alpha_l}{2}\right)\sum_{i=n_1+1,i\ne l}^{n}\alpha_i -{\rm I}(l \in R_2) \alpha_l\sum_{j=1, j\ne l}^{n}\left((1-\alpha_{j})\widetilde{p}+\frac{\alpha_{j}}{2}\right)\right].
\end{align*}
\end{theorem}

A test for the hypothesis $H_0:p=1/2$, based on the  test statistic
$$\widetilde{Z} = \frac{\widetilde{p} - \frac{1}{2}}{\sqrt{\widehat{{\rm Var}(\widetilde{p})}}}$$  would reject $H_0$ when $|\widetilde Z|$ exceeds the upper $\alpha/2$ quantile of the standard normal distribution, i.e. $|\widetilde{Z}|>z_{1-\alpha/2}$.  Further,
a $(1-\alpha)\%$ confidence interval of $p$ can be obtained from   $$\bigg(\widetilde{p}-z_{1-\frac{\alpha}{2}}\sqrt{\widehat{{\rm Var}(\widetilde{p})}}, \widetilde{p}+z_{1-\frac{\alpha}{2}}\sqrt{\widehat{{\rm Var}(\widetilde{p})}}\bigg).$$  

\subsection{Finite Sample Approximation}
 
The normal approximation might be inaccurate for small samples since ideally both number of complete and incomplete clusters need to be large  for the asymptotics to apply.  Satterthwaite-Smith-Welch (SSW) type approximation using a Student's $t$ distribution is a popular approach for small-sample approximation, and its idea could be generalized to work in the nonparametric situation as well \citep{BDM1997, BM2000, BKPP2017, CKH2021}).  Denoting $\widehat{V_i} = (\widehat{W}_i - \widehat{E(W_i)})^2$, an estimate of the degrees of freedom can be computed analogous to SSW as 
$$\widehat{\nu} = \frac{(\sum_{i=1}^{n} \widehat{V_i})^2}{\frac{(\sum_{i=1}^{n_1} \widehat{V_i})^2}{n_1-1} + \frac{(\sum_{i=n_1+1}^{n_1+n_2} \widehat{V_i})^2}{n_2-1} + \frac{(\sum_{i=n_1+n_2 + 1}^{n} \widehat{V_i})^2}{n_c-1}}$$
  It can be seen that $\widehat{\nu}\rightarrow\infty$ as $n\rightarrow\infty$, meaning that as the number of clusters gets larger the $t-$approximation converges to a standard normal distribution, which is anticipated asymptotically.  This small-sample approximation works well for both small number of clusters and small cluster sizes, provided that $n_1$, $n_2$ and $n_c$ are all larger than 1.  In the case when one or more of $n_1$, $n_2$ and $n_c$ is $0$ or $1$, the corresponding denominator can be replaced by $1$ in order to obtain an approximation; however, a simple approximation like the above is not expected to work accurately under  the extreme cases.    
  
  A $(1-\alpha)\%$ small-sample confidence interval of $p$ can be expressed by swapping the normal percentiles to the corresponding $t$ percentiles:
$$\bigg(\widetilde{p}-t^*_{1-\frac{\alpha}{2}}\sqrt{\widehat{{\rm Var}(\widetilde{p})}}, \widetilde{p}+t^*_{1-\frac{\alpha}{2}}\sqrt{\widehat{{\rm Var}(\widetilde{p})}}\bigg)$$
where $t_{1-\frac{\alpha}{2}}^*$ is the $(1-\frac{\alpha}{2})$ percentile of the  $t_{\widehat{\nu}}$ distribution. 

\subsection{Hoffman-type Variance Estimator}\label{sec:hoffvar}
Another method for estimating $Var(\widetilde{p})$ inspired by \citet{HSW2001} is based on the properties of conditional variance. Specifically,
{\begin{align*}
    \widehat{{\rm Var}_H(\widetilde{p})} = \frac{1}{[E(m_1^*m_2^*|\bm{X,g})]^2}\widehat{{\rm Var}[E(U^*|\bm{X,g})]},
\end{align*}
where the estimated variance term on the right hand side can be decomposed into two pieces
\begin{align*}
    \widehat{{\rm Var}[E(U^*|\bm{X,g})]}= \widehat{E[{\rm Var}(U^*|\bm{X^*,g^*})] }- \widehat{E[{\rm Var}(U^*|\bm{X,g})]},
\end{align*}
and each piece can be estimated by Monte Carlo method.  For the first piece, we take the average of estimated variances calculated following the variance estimator for two independent samples as in \citet{BM2000} with a large number of  within-cluster resamples.  For the second piece, we calculate the average squared error of $U^*$ based on a large number of {resamples}.  However, the resulting estimator is a difference between two quantities, and an issue might occasionally occur that the variance estimate would be negative, rendering the test result unavailable.  In spite of this, we propose an alternative test based on this new variance estimator with the following test statistic:
\begin{align*}
    \widetilde{Z}_{H} = \frac{\widetilde{p} - \frac{1}{2}}{\sqrt{\widehat{Var}_H(\widetilde{p})}}.
\end{align*}
It should be mentioned that $\widetilde{Z}$ is generally preferred over $\widetilde{Z}_{H}$ since $\widetilde{Z}_{H}$ is a Monte Carlo estimator that the realized value of $\widehat{{\rm Var}_H(\widetilde{p})}$ can occasionally be negative.  Nevertheless,  $\widetilde{Z}_H$ can serve as an alternative option when it is available.  

\subsection{Other Test Statistics}\label{subsec:otherts}
We have introduced our main test statistic $\widetilde{Z}$ together with a small-sample approximation, as well as an alternative test statistic $\widetilde{Z}_{H}$.  However, these two test statistics are both based on the estimator $\widetilde{p}$, and it might be worthwhile investigating on potential test{s} that are based on $\widehat{p}^*$ or $\widehat{p}$.

Based on $\widehat{p}^*$, we propose the test statistic
\begin{align*}
    \widehat{Z}^* = \frac{\widehat{p}^*-\frac{1}{2}}{\widehat{\sigma}^{*}},
\end{align*}
where $\widehat{\sigma}^{*2}$ is the variance estimator of the Mann-Whitney effect estimator for two independent samples as  in \citet{BM2000} calculated from a single {within-cluster resample}.
Clearly, $\widehat{Z}^*$ is a naive test statistic that is not data-efficient.  
To overcome this problem, we propose a  test  based on $\widehat{p}$
\begin{align*}
    \widehat{Z} = \frac{\widehat p - \frac{1}{2}}{\sqrt{\widehat{{\rm Var}(\widehat{p})}}},
\end{align*}
where $\widehat{{\rm Var}(\widehat{p})}$ is a Monte Carlo variance estimator calculated in a very similar manner as $\widehat{{\rm Var}_{H}(\widetilde{p})}$ based on a large number of  {resamples}. {However, }the test statistic  $\widehat{Z}$ bears the issue that its variance estimator can occasionally be negative, just like $\widetilde{Z}_H$.



\section{Simulation Study}\label{sec:sim}
The objective of the simulation study is to investigate the finite sample performance of the proposed tests and confidence intervals in terms of type {\rm I} error rate, power and coverage probability.  We will study the effects {of} data distribution, number of clusters, cluster sizes, intra- and inter-cluster correlation, and group variance.  We compare the performance of our methods with some existing in the literature both under informative and noninformative cluster size set-ups.  The tests that are of the most interest are tests proposed by \citet{DS2005} (abbreviated as DS hereinafter), and \citet{CKH2021} (CKH).  We denote $\widetilde{T}$ as the small-sample approximation counterpart of $\widetilde{Z}$.  The small-sample approximation of \citet{CKH2021} is also included (abbreviated as $\text{CKH}_{T}$).  Hoffman-type tests based on test statistics $\widehat{Z}$ and $\widetilde{Z}_{H}$, {as well as the naive test $\widehat{Z}^*$} are also included for comparisons.  Monte Carlo calculation of $\widehat{Z}$ and $\widetilde{Z}_{H}$ are based on $1,000$ repetitions. In the exploratory stages we also tried $10,000$ repetitions for these calculations, but results were not seen to differ much.  Other tests such as \citet{RGL2004} and \citet{LHNO2010} are excluded here because it has been shown by \citet{DS2005} that \citet{RGL2004} performs worse in terms of power under the setting of informative cluster size, and simulations from \citet{CKH2021} concluded that the test proposed by \citet{LHNO2010} is too conservative in terms of achieved type I error rate.

\subsection{Simulation Setup}\label{subsec:simsetup}
 For distribution of the data, we consider multivariate Gaussian and multivariate Cauchy distributions.  A discretized version of multivariate Gaussian distribution was also used in exploratory stages, but results are not seen much different from that of the {regular} multivariate Gaussian.  For number of clusters, we consider the cases when $n_1$ is larger and when $n_c$ is larger.  Two settings are considered for the intra-cluster group size, namely $m_i^{(j)}-1$ is drawn independently from  ${\rm Binomial}(n=2, p=0.3)$ or a ${\rm Binomial}(n=9, p=0.3)$  distribution.  We consider homoskedastic as well as heteroskedastic situations for group variances.  Note that under the heteroskedastic situation $F_1\ne F_2$ but the null hypothesis $H_0:p=1/2$ holds.  For within- and between-cluster correlations, the complete clusters data are generated with the following variance-covariance structure (in cases of multivariate Cauchy distribution, this refers to the scale matrix)
\[\bm{\Sigma} = \begin{pmatrix}
\sigma_1^2\bm{I}_{m_i^{(1)}}+\rho_1\sigma_1^2\bm{J}_{m_i^{(1)}\times m_i^{(1)}} & \rho_{12}\sigma_1\sigma_2 \bm{J}_{m_i^{(1)}\times m_i^{(2)}}\\
    \rho_{12}\sigma_1\sigma_2 \bm{J}_{m_i^{(1)}\times m_i^{(2)}} & \sigma_2^2\bm{I}_{m_i^{(2)}}+\rho_2\sigma_2^2\bm{J}_{m_i^{(2)}\times m_i^{(2)}}
\end{pmatrix}
\]
where $\bm{I}_m$ is the $m\times m$ identity matrix, $\bm{J}_{m\times n}$ is the $m\times n$ matrix where all entries are 1.  This way $\rho_1$ and $\rho_2$ are the parameter for intra-cluster correlations for the two types of incomplete clusters, and $\rho_{12}$ is the parameter for inter-cluster correlations.  Moreover, homoskedasticity and heteroskedasticity can be altered by setting $\sigma_1$ and $\sigma_2$ to be either equal or unequal.

All results are based on 10,000 simulation replications.  The nominal type I error rate and confidence coefficient are always set at $0.05$ and $0.95$, respectively.

\subsection{Type I Error Rate}\label{sec:at1eres}
Tables \ref{table2} and \ref{table3} show the actual type I error rate (\%) when both groups follow a multivariate normal distribution.  All the tests included here have reasonable performance.  For number of clusters, it seems that all tests have slightly better performances when the clusters are more balanced ({i.e.,} when $n_1 = n_2$ with a large $n_c$).  This is expected since balanced data provides {more} accurate estimation of both $F_1$ and $F_2$.  For variances, it appears that unequal variances lead to slightly worse performances for all tests.  On the other hand, different combinations of $(\rho_1,\rho_2,\rho_{12})$ are not seen to have a noticeable impact; however, theoretically we do prefer all of these parameters to be small.  For intra-cluster group sizes, we also do not see an obvious effect.

Comparing the tests, the small-sample approximation ($\widetilde{T}$) seems to have the most correct achieved type I error rate, which is anticipated since the expected  {intra-cluster group} size is small (1.6 when $\max(m_i^{(j)})=3$, and 3.7 when $\max(m_i^{(j)})=10$), and the total number of clusters {(40)} is not very big.  To be more specific, the estimated degrees of freedom are only between 15 to 28.  Additionally, performances of the Hoffman-type tests $\widehat{Z}$ and $\widetilde{Z}_H$ are quite comparable to $\widetilde{Z}$.  Surprisingly, the performance of the naive $\widehat{Z}^*$ is not so bad on average when compared to $\widehat{Z}$, even though its calculation is only based on one chosen resample.  That said, the test $\widehat{Z}$ is seen to slightly outperform $\widehat{Z}^*$ under most cases.  Also, $\widetilde{Z}_H$ appears to have the most inflated type I error rate, especially when the design is more unbalanced (i.e. in the case of $n_1 > n_c$).  We will see also in later simulations that $\widetilde{Z}_{H}$ is liberal.  In this sense, $\widetilde{Z}$ is evidently preferred over $\widetilde{Z}_H$.  On the other hand, DS is performing very well when $\sigma_1^2 = \sigma_2^2$; however, its performance is seen to be particularly affected by heteroskedasticity.  CKH {is} seen comparable with $\widetilde{Z}$.  In addition, $\widetilde{Z}_H$ can occasionally have negative variance issue (around 1 to 2 times on average per 10,000 simulations) as mentioned in Section \ref{sec:hoffvar}, making the test less favorable.  In theory, both $\widehat{Z}$ and $\widetilde{Z}_H$ can have this issue, but  {practically} speaking $\widetilde{Z}_H$ has a higher chance of this occurring.  For the small-sample approximation, both $\widetilde{T}$ and $\text{CKH}_T$ appear to do a better job than their original counterpart, because of the small estimated degrees of freedom as mentioned earlier.  To sum up, our small-sample approximation test statistic $\widetilde{T}$, together with $\text{CKH}_{T}$ work the best, and it is difficult to tell which one is better based on just Tables \ref{table2} and \ref{table3}.  DS only works well under homoskedasticity.

\begin{table}[H]
	\begin{center}
		\begin{tabular}{||c|c|c|c|c|c|c|c|c|c||} 
			\hline
			\multicolumn{10}{||c||}{$(n_1,n_2,n_c) = (20,10,10)$,  max$(m_i^{(j)})$ = 2}\\
			\hline
			$(\rho_1, \rho_2, \rho_{12})$ & $(\sigma_1^2, \sigma_2^2)$ & $\widetilde{Z}$ & $\widetilde{T}$ & $\widehat{Z}^*$ & $\widehat{Z}$ & $\widetilde{Z}_{H}$ & DS & CKH & $\text{CKH}_{T}$\\[0.2ex]
			\hline
			\multirow{2}{*}{(0.9,0.9,0.1)} & (1,1) & 6.09 & 5.25 & 6.47 & 5.88 & 6.91 & 5.13 & 6.10 & 5.11\\
			& (1,5) & 6.86 & 5.23 & 6.90 & 6.85 & 7.38 & 6.44 & 6.52 & 5.20\\
			\hline
			\multirow{2}{*}{(0.1,0.9,0.9)} & (1,1) & 6.06 & 5.08 & 6.19 & 6.03 & 7.11 & 5.20 & 5.70 & 4.75\\
			& (1,5) & 6.73 & 5.18 & 6.81 & 6.64 & 7.32 & 7.31 & 6.23 & 4.90\\
			\hline
			\multirow{2}{*}{(0.1,0.1,0.9)} & (1,1) & 6.14 & 4.93 & 6.59 & 5.90 & 7.19 & 5.29 & 5.49 & 4.54\\
			& (1,5) & 6.78 & 5.18 & 6.97 & 6.69 & 7.83 & 7.02 & 5.62 & 4.34\\
			\hline
			\multicolumn{10}{||c||}{$(n_1,n_2,n_c) = (20,10,10)$, max$(m_i^{(j)})$ = 9}\\
			\hline
			$(\rho_1, \rho_2, \rho_{12})$ & $(\sigma_1^2, \sigma_2^2)$ & $\widetilde{Z}$ & $\widetilde{T}$ & $\widehat{Z}^*$ & $\widehat{Z}$ & $\widetilde{Z}_{H}$ & DS & CKH & $\text{CKH}_{T}$\\[0.2ex]
			\hline
			\multirow{2}{*}{(0.9,0.9,0.1)} & (1,1) & 6.73 & 5.74 & 6.48 & 6.45 & 7.29 & 5.44 & 6.32 & 5.15\\
			& (1,5) & 7.51 & 5.80 & 7.16 & 7.30 & 7.97 & 6.97 & 6.69 & 5.37\\
			\hline
			\multirow{2}{*}{(0.1,0.9,0.9)} & (1,1) & 6.44 & 5.25 & 6.20 & 6.35 & 7.84 & 5.81 & 5.95 & 4.80\\
			& (1,5) & 7.17 & 5.39 & 6.91 & 6.99 & 7.86 & 8.11 & 6.14 & 4.73\\
			\hline
			\multirow{2}{*}{(0.1,0.1,0.9)} & (1,1) & 6.17 & 5.29 & 6.42 & 5.49 & 8.15* & 5.66 & 5.73 & 4.70\\
			& (1,5) & 6.89 & 5.21 & 6.87 & 6.23 & 8.69* & 6.73 & 6.03 & 4.74\\
			\hline
		\end{tabular}
		{\caption{\label{table2}\small Empirical type I error rate (\%) under ignorable cluster size when both groups follow multivariate Gaussian distributions and nominal type I error rate is set to be 5\%. $n_1$, $n_2$ are the number of incomplete clusters of group 1, and 2, respectively.  $n_c$ is the number of complete clusters.  $\max(m_i^{(j)})$ stands for the maximum allowed intra-cluster group (ICG) sizes.  $\sigma_1^2$ and $\sigma_2^2$ denote the theoretical variances of group 1 and 2 observations, respectively.  $\rho_1$ and $\rho_2$ denote the intra-cluster correlations within group 1 and 2, respectively.  $\rho_{12}$ denotes the inter-cluster correlation.  These parameters are directly associated with the theoretical variance-covariance matrix of the distributions, as mentioned in Section \ref{subsec:simsetup}.  The methods $\widetilde{Z}$, $\widetilde{T}$, $\widehat{Z}^*$,  $\widehat{Z}$ and $\widetilde{Z}_{H}$ are the proposed method and its variants.  DS is the test proposed in \cite{DS2005} whereas CKH and ${\rm CKH}_{T}$ are the test by \citet{CKH2021} and its small-sample approximation.  Design here is unbalanced with slightly more incomplete clusters in group 1.  The superscript $*$ on the numbers in the body of the table indicates at least 1 (out of 10,000 simulations) result is not available due to negative variance estimation.}}
	\end{center}
\end{table}

\begin{table}[H]
	\begin{center}
		\begin{tabular}{||c|c|c|c|c|c|c|c|c|c||} 
			\hline
			\multicolumn{10}{||c||}{$(n_1,n_2,n_c) = (10,10,20)$, max$(m_i^{(j)})$ = 2}\\
			\hline
			$(\rho_1, \rho_2, \rho_{12})$ & $(\sigma_1^2, \sigma_2^2)$ & $\widetilde{Z}$ & $\widetilde{T}$ & $\widehat{Z}^*$ & $\widehat{Z}$ & $\widetilde{Z}_{H}$ & DS & CKH & $\text{CKH}_{T}$\\[0.2ex]
			\hline
			\multirow{2}{*}{(0.9,0.9,0.1)} & (1,1) & 5.94 & 5.20 & 5.97 & 5.78 & 5.97 & 5.07 & 5.55 & 4.71\\
			& (1,5) & 6.09 & 4.94 & 6.22 & 6.05 & 6.29 & 6.32 & 5.75 & 4.92\\
			\hline
			\multirow{2}{*}{(0.1,0.9,0.9)} & (1,1) & 5.61 & 4.47 & 6.14 & 5.78 & 6.09 & 5.25 & 5.11 & 4.35\\
			& (1,5) & 5.78 & 4.53 & 6.18 & 6.15 & 6.33 & 7.68 & 5.70 & 4.68\\
			\hline
			\multirow{2}{*}{(0.1,0.1,0.9)} & (1,1) & 5.77 & 4.60 & 6.06 & 5.52 & 5.88 & 5.35 & 4.92 & 4.12\\
			& (1,5) & 5.79 & 4.57 & 6.28 & 6.09 & 6.30 & 7.24 & 5.53 & 4.75\\
			\hline
			\multicolumn{10}{||c||}{$(n_1,n_2,n_c) = (10,10,20)$, max$(m_i^{(j)})$ = 9}\\
			\hline
			$(\rho_1, \rho_2, \rho_{12})$ & $(\sigma_1^2, \sigma_2^2)$ & $\widetilde{Z}$ & $\widetilde{T}$ & $\widehat{Z}^*$ & $\widehat{Z}$ & $\widetilde{Z}_{H}$ & DS & CKH & $\text{CKH}_{T}$\\[0.2ex]
			\hline
			\multirow{2}{*}{(0.9,0.9,0.1)} & (1,1) & 5.90 & 5.00 & 6.22 & 5.52 & 5.80 & 4.91 & 5.97 & 4.93\\
			& (1,5) & 5.74 & 4.63 & 6.08 & 5.70 & 5.96 & 5.91 & 5.95 & 4.97\\
			\hline
			\multirow{2}{*}{(0.1,0.9,0.9)} & (1,1) & 5.21 & 4.13 & 6.48 & 5.22 & 5.55 & 5.03 & 5.41 & 4.55\\
			& (1,5) & 5.67 & 4.29 & 6.13 & 5.87 & 6.25 & 7.98 & 5.29 & 4.31\\
			\hline
			\multirow{2}{*}{(0.1,0.1,0.9)} & (1,1) & 6.05 & 5.21 & 6.46 & 5.28 & 5.81 & 5.70 & 5.97 & 5.23\\
			& (1,5) & 5.65 & 4.48 & 6.21 & 5.46 & 5.87 & 6.07 & 5.64 & 4.71\\
			\hline
		\end{tabular}\\
		{{\caption{\small\label{table3} Empirical type I error rate (\%) under ignorable cluster size when both groups follow multivariate Gaussian distributions and nominal type I error rate is set to be 5\%. $n_1$, $n_2$ are the number of incomplete clusters of group 1, and 2, respectively.  $n_c$ is the number of complete clusters.  $\max(m_i^{(j)})$ stands for the maximum allowed intra-cluster group (ICG) sizes.  The parameters $\sigma_1^2$, $\sigma_2^2$, $\rho_1$, $\rho_2$ and $\rho_{12}$ are directly associated with the theoretical variance-covariance matrix of the distributions, as mentioned in Section \ref{subsec:simsetup}.  The methods $\widetilde{Z}$, $\widetilde{T}$, $\widehat{Z}^*$, $\widehat{Z}$ and $\widetilde{Z}_{H}$ are the proposed method and its variants.  DS is the test proposed in \cite{DS2005} whereas CKH and ${\rm CKH}_{T}$ are the test by  \citet{CKH2021} and its small-sample approximation.  Design here is comparatively balanced with more complete clusters.}}}
	\end{center}
\end{table}

For Tables S1 and S2 in the supplementary material, the actual type I error rate (\%) are shown for the setting where both groups follow multivariate Cauchy distribution.  Generally, we can draw a very similar conclusion as with Tables \ref{table2} and \ref{table3}, meaning that the underlying distribution does not have a big impact on the results.  Particularly, all tests have a worse achieved type I error rate when the scale parameters $\sigma_1$ and $\sigma_2$ differ, and when the design is less balanced.  The combination of parameters $(\rho_1,\rho_2,\rho_{12})$, together with maximum intra-cluster group sizes are still not seen to be affecting the results much.  $\widetilde{Z}_H$ is still the most liberal test, and the two small-sample approximation tests still give the most correct results.  Notably, there was only one occurrence of negative variance estimation in the 10,000 replications for $\widehat{Z}$ and this is {still} less frequent compared to $\widetilde{Z}_{H}$.

 To illustrate the effect of unequal variances under the null hypothesis, Table S3 in the supplementary material shows the actual type I error rate (\%) when both groups follow a multivariate normal distribution but $\sigma_2$ getting more and more different from $\sigma_1$.  Noticeably DS has the most inflated achieved type I error rate compared to all other tests as $\sigma_2$ gets large, from around $0.05$ to $0.12$ when maximum {intra-cluster group} size is 3, and from around 0.05 to 0.10 when maximum intra-cluster group size is 10.  While all other tests may also have a slightly inflated achieved type I error rate when $\sigma_2 = 10$ compared to that of $\sigma_2 = 1$, none of these effects are as obvious as DS.  That said, we know that DS is intended to  detect departure from its null hypothesis of $F_1 = F_2$, so the inflated type I error rate is not unreasonable.  On the other hand, CKH performs very well in ignoring $F_1\ne F_2$ by maintaining an achieved type I error rate of around 0.05 when $\sigma_2$ gets large, and $\widetilde{Z}$ seems to perform better than the Hoffman variants $\widehat{Z}$ and $\widetilde{Z}_H$, and the naive test $\widehat{Z}^*$, not mentioning that both small-sample approximations $\widetilde{T}$ and $\text{CKH}_T$ still perform very well, which echoes our previous findings.  Both $\widehat{Z}$ and $\widetilde{Z}_{H}$ had one to two negative variances per 10,000 simulations when maximum intra-cluster group size is 10.

 Figure \ref{fig:unequalvar} plots the achieved type I error rate against the ratio of variances in the two groups.  The dotted red horizontal line shows the nominal type I error rate of $0.05$, whereas solid curves in different colors shows the achieved type I error rate of the different tests.  It is clear that DS appears well above all other curves, meaning that it has the highest achieved type I error rate among all tests in both plots.  In the top plot, we see that the two small-sample approximations have the most correct achieved type I error rate.  In the bottom plot, both small-sample approximations are still good, with CKH also performing reasonably well.  We can also see in both plots that $\widetilde{Z}$ has a better achieved type I error rate compared with its Hoffman counterparts $\widehat{Z}$, $\widetilde{Z}_H$, and the naive test $\widehat{Z}^*$.   These simulation results indeed show that  our tests have better performances compared to DS under this heteroskedastic setting.
 
 \begin{figure}[H]
 	\centering
 	\includegraphics[scale = 0.8]{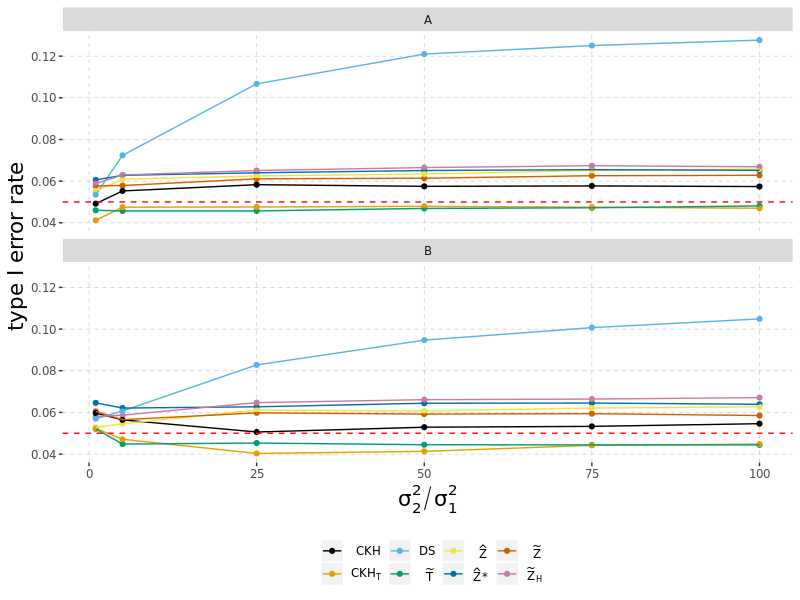}
 	{\caption{\label{fig:unequalvar} Plot of achieved type I error rate by ratio of theoretical variances of the two groups for all 8 tests.  Panel A shows maximum allowed ICG size 2, and panel B shows maximum allowed ICG size 9.  This plot serves as a visualization of Table S3.  Larger ratio provides stronger evidence of $F_1\ne F_2$, under which cases DS does not seem to perform well.}}
 \end{figure}

\subsection{Power}\label{subsec:poweranalysis}

The previous simulations were in the noninformative cluster size set-up.  We now consider power analysis under the setting as described in Section \ref{sec:motivation} where cluster size is informative.  For complete clusters, we set the {intra-cluster group sizes of the} clusters  {to be} equal, i.e., $m_i^{(1)} = m_i^{(2)}$ drawn randomly from  $\{c_1,c_2\}$.  Recall that cluster size is related to the outcome except when $c_1 = c_2$.  However, notice that if $|c_1-c_2|$ is too large, then we would end up with significant number of observations from a certain group being different from the other group, compromising the information provided through cluster size.  That is, a test would not need the assumption of informative cluster size in order to detect a difference if most observations from group 1 is smaller than those from group 2.  This is also reflected in Table \ref{table1} where $p_0$ can be seen noticeably different from 0.5 as the absolute difference in $c_1$ and $c_2$ increases to 5.

Tables \ref{table7} and \ref{table8} are extensions of Table \ref{table1} that show the power (\%) of each test in detecting $p\ne\frac{1}{2}$ for different values of $p$ where each cluster is generated from multivariate normal distributions with potentially different means depending on the preset values of $c_1$ and $c_2$ as mentioned in Section \ref{sec:motivation}, fixing the combination of $(n_1, n_2, n_c)$ to be $(20,10,10)$, and with varying inter- and intra-cluster correlations.  As we have seen in Figure \ref{fig:introcomparison}, here we can clearly see again that both CKH and its small-sample approximation are built for testing $p_0 = \frac{1}{2}$ whereas the other tests appear to be more sensitive for testing $p=\frac{1}{2}$.  We can see from the results that as $p$ goes from 0.5 up to 0.74 or down to 0.26, the power of all tests increased except for CKH, since $p_0$ only went from 0.5 up to 0.52 or down to 0.48 under these situations.  However, as the difference keeps increasing, CKH is gradually gaining power.  For instance, when $c_1 = 2$ and $c_2 = 7$, $p_0$ began to show a noticeable difference from 0.5, and this is because about $\frac{7}{9}$ or $77.8\%$ of observations from group 1 are generally larger than about 77.8\% of observations from group 2 under this situation. This in turn undermines the null hypothesis of $p_0 = \frac{1}{2}$ under ignorable cluster size, eventually leading to an increase in the power of CKH. On the other hand, DS is giving a consistent result here since the fact that $p$ gets more and more different from $\frac{1}{2}$ implies that $F_1$ is increasingly different from $F_2$. Additionally, when the null hypothesis is violated, DS can be seen less powerful here than the five tests that we proposed, and this balances out with its previous outstanding performance in the achieved type I error rate (since this shows that the test is conservative).  In contrast, $\widetilde{T}$ is doing a better job by both achieving a small type I error rate when $H_0$ holds, and a good power when $H_0$ does not hold.  It can be further noticed here that $\widehat{Z}^*$ is slightly less powerful than $\widehat{Z}$, which is in accordance with our anticipation.  Also, there seems to be a trend that $\widetilde{Z}_{H}$ has negative variance estimates more frequently when $p$ gets large, this may be due to the nature of how $\widetilde{Z}_{H}$ is constructed.

\begin{table}[H]
	\centering
	\begin{tabular}{||c|c|c|c|c|c|c|c|c|c|c|c|c||}
		\hline
		\multicolumn{13}{||c||}{$(n_1,n_2,n_c) = (20,10,10)$}\\
		\multicolumn{13}{||c||}{$(\sigma_1^2, \sigma_2^2, \rho_1, \rho_2, \rho_{12}) = (1,1,0.1,0.1,0.9)$}\\
		\hline
		$c_1$ & $c_2$ & $\mu_d$ & $p$ & $p_0$ & $\widetilde{Z}$ & $\widetilde{T}$ & $\widehat{Z}^*$ & $\widehat{Z}$ & $\widetilde{Z}_{H}$ & DS & CKH & $\text{CKH}_{T}$\\[0.2ex]
		\hline
		2 & 2 & 0 & 0.50 & 0.50 & 6.08 & 4.99 & 6.40 & 5.46 & 7.76 & 5.35 & 5.15 & 4.28\\
		\hline
		2 & 3 & -1 & 0.63 & 0.53 & 34.93 & 31.95 & 30.02 & 33.50 & 37.59 & 27.77 & 6.26 & 5.05\\
		\hline
		3 & 2 & 1 & 0.37 & 0.47 & 34.10 & 31.28 & 29.07 & 32.85 & 33.94 & 27.20 & 6.27 & 5.21\\
		\hline
		2 & 4 & -2 & 0.71 & 0.52 & 68.45 & 65.69 & 63.82 & 67.23 & 70.50* & 57.77 & 6.53 & 5.26\\
		\hline
		4 & 2 & 2 & 0.29 & 0.48 & 67.72 & 65.15 & 63.45 & 66.38 & 67.26 & 57.83 & 6.02 & 4.94\\
		\hline
		2 & 5 & -3 & 0.74 & 0.48 & 79.28 & 77.15 & 76.84 & 77.90 & 80.51* & 69.22 & 5.98 & 4.76\\
		\hline
		5 & 2 & 3 & 0.26 & 0.52 & 78.65 & 76.38 & 76.15 & 77.43 & 77.89 & 68.32 & 5.90 & 4.74\\
		\hline
		2 & 6 & -4 & 0.75 & 0.44 & 81.09 & 79.69 & 79.14 & 80.78 & 81.72* & 70.28 & 11.01 & 9.49\\
		\hline
		6 & 2 & 4 & 0.25 & 0.56 & 81.03 & 79.65 & 79.09 & 80.70 & 80.97 & 70.43 & 11.06 & 9.43\\
		\hline
		2 & 7 & -5 & 0.75 & 0.40 & 80.86 & 79.55 & 79.26 & 80.53 & 81.62* & 68.88 & 21.72 & 17.69\\
		\hline
		7 & 2 & 5 & 0.25 & 0.60 & 82.41 & 81.07 & 80.57 & 82.16 & 82.62 & 69.88 & 20.78 & 17.02\\
		\hline
	\end{tabular}\\
	{ {\caption{\small\label{table7} Achieved power (\%) under informative cluster size for detecting a violation of $p=\frac{1}{2}$ when both groups follow multivariate Gaussian distributions. $n_1$, $n_2$ are the number of incomplete clusters of group 1, and 2, respectively.  $n_c$ is the number of complete clusters.  The parameters $\sigma_1^2$, $\sigma_2^2$, $\rho_1$, $\rho_2$ and $\rho_{12}$ are directly associated with the theoretical variance-covariance matrix of the distributions, as mentioned in Section \ref{subsec:simsetup}.  $p$ and $p_0$ are the theoretical values of the WMW treatment effect under informative ($p$) and ignorable cluster size ($p_0$), respectively.  The parameters $c_1$ and $c_2$ control the informativeness of the cluster size.  The case $c_1=c_2$ is noninformative (igorable) cluster size.   $\mu_d$ is the difference in theoretical means between the two groups.  The methods $\widetilde{Z}$, $\widetilde{T}$, $\widehat{Z}^*$, $\widehat{Z}$ and $\widetilde{Z}_{H}$ are the proposed method and its variants.  DS is the test proposed in \cite{DS2005} whereas CKH and ${\rm CKH}_{T}$ are the test by \citet{CKH2021} and its small-sample approximation.  The superscript $*$ on the numbers in the body of the table indicates at least 1 (out of 10,000 simulations) result is not available due to negative variance estimation.}}}
\end{table}

\begin{table}[H]
	\begin{center}
		\begin{tabular}{||c|c|c|c|c|c|c|c|c|c|c|c|c||}
			\hline
			\multicolumn{13}{||c||}{$(n_1,n_2,n_c) = (20,10,10)$}\\
			\multicolumn{13}{||c||}{$(\sigma_1^2, \sigma_2^2, \rho_1, \rho_2, \rho_{12}) = (1,1,0.9,0.9,0.1)$}\\
			\hline
			$c_1$ & $c_2$ & $\mu_d$ & $p$ & $p_0$ & $\widetilde{Z}$ & $\widetilde{T}$ & $\widehat{Z}^*$ & $\widehat{Z}$ & $\widetilde{Z}_{H}$ & DS & CKH & $\text{CKH}_{T}$\\[0.2ex]
			\hline
			2 & 2 & 0 & 0.50 & 0.50 & 6.18 & 5.09 & 6.19 & 5.85 & 6.72 & 5.02 & 5.47 & 4.49\\
			\hline
			2 & 3 & -1 & 0.63 & 0.53 & 34.23 & 31.42 & 31.98 & 33.23 & 36.65* & 27.16 & 6.54 & 5.38\\
			\hline
			3 & 2 & 1 & 0.37 & 0.47 & 33.59 & 30.72 & 30.95 & 32.62 & 33.61 & 26.66 & 6.65 & 5.57\\
			\hline
			2 & 4 & -2 & 0.71 & 0.52 & 68.87 & 66.18 & 66.51 & 67.68 & 70.94* & 57.42 & 6.44 & 5.40\\
			\hline
			4 & 2 & 2 & 0.29 & 0.48 & 68.20 & 65.68 & 65.57 & 67.13 & 67.81 & 57.21 & 6.40 & 5.16\\
			\hline
			2 & 5 & -3 & 0.74 & 0.48 & 79.67 & 77.48 & 77.89 & 78.62 & 80.79* & 68.03 & 6.04 & 4.67\\
			\hline
			5 & 2 & 3 & 0.26 & 0.52 & 78.85 & 76.74 & 76.83 & 78.15 & 78.49 & 67.24 & 5.72 & 4.69\\
			\hline
			2 & 6 & -4 & 0.75 & 0.44 & 80.97 & 79.41 & 79.18 & 80.72 & 81.73* & 68.62 & 11.03 & 9.54\\
			\hline
			6 & 2 & 4 & 0.25 & 0.56 & 80.88 & 79.54 & 79.04 & 80.66 & 80.87 & 68.50 & 11.14 & 9.68\\
			\hline
			2 & 7 & -5 & 0.75 & 0.40 & 80.68 & 79.29 & 78.91 & 80.46 & 81.58* & 65.44 & 22.37 & 18.03\\
			\hline
			7 & 2 & 5 & 0.25 & 0.60 & 82.22 & 80.95 & 80.53 & 82.07 & 82.28 & 66.72 & 21.39 & 17.39\\
			\hline
		\end{tabular}\\
		{ {\caption{\small\label{table8} Achieved power (\%) under informative cluster size for detecting a violation of $p=\frac{1}{2}$ when both groups follow multivariate Gaussian distributions. $n_1$, $n_2$ are the number of incomplete clusters of group 1, and 2, respectively.  $n_c$ is the number of complete clusters.  The parameters $\sigma_1^2$, $\sigma_2^2$, $\rho_1$, $\rho_2$ and $\rho_{12}$ are directly associated with the theoretical variance-covariance matrix of the distributions, as mentioned in Section \ref{subsec:simsetup}.  $p$ and $p_0$ are the theoretical values of the WMW treatment effect under informative ($p$) and ignorable cluster size ($p_0$), respectively.  The parameters $c_1$ and $c_2$ control the informativeness of the cluster size.  The case $c_1=c_2$ is noninformative (igorable) cluster size.   $\mu_d$ is the difference in theoretical means between the two groups.  The methods $\widetilde{Z}$, $\widetilde{T}$, $\widehat{Z}^*$, $\widehat{Z}$ and $\widetilde{Z}_{H}$ are the proposed method and its variants.  DS is the test proposed in \cite{DS2005} whereas CKH and ${\rm CKH}_{T}$ are the test by \citet{CKH2021} and its small-sample approximation.  The superscript $*$ on the numbers in the body of the table indicates at least 1 (out of 10,000 simulations) result is not available due to negative variance estimation.  In particular, simulation results from this table uses a slightly different variance-covariance matrix compared to Table \ref{table7}.}}}
	\end{center}
\end{table}

For Tables S4 and S5 in the supplementary material, simulation results similar to Tables \ref{table7} and \ref{table8} are presented when the underlying distribution is a multivariate Cauchy, where {$c_1$} and {$c_2$ here} are location parameters {instead of means}.  The results and conclusions here are very similar to Tables \ref{table7} and \ref{table8}: all tests are consistently gaining power except for CKH, and finally CKH gained a little power in the case when $p_0$ is up to 0.56 or down to 0.44.  The test $\widetilde{Z}_H$ is still seen to have more negative variance estimations when $p$ is large, and DS still have lower power compared with $\widetilde{Z}$, $\widetilde{T}$, $\widehat{Z}$, and $\widetilde{Z}_H$.  More obviously, $\widehat{Z}^*$ is seen much less powerful than $\widehat{Z}$ compared to results in Tables \ref{table7} and \ref{table8}.

A plot of the achieved power against $p$ is shown in Figure \ref{fig:powerICS}.  It can be clearly seen that CKH is ``outperformed'' by other tests in all scenarios due to the fact that it is ignoring informative cluster sizes and the test is built for testing $p_0=\frac{1}{2}$.  We may also notice that $\widetilde{Z}_H$ tends to be rejecting the null hypothesis the most often, and DS tends to be rejecting the null hypothesis the least often.  Furthermore, the performances of our tests $\widetilde{Z}$, $\widetilde{T}$, $\widehat{Z}$ and $\widetilde{Z}_H$ are very similar in all four plots, and $\widehat{Z}^*$ can be seen less powerful in the bottom-left plot where more variability is present.  The reason why DS is comparatively less powerful here might be because that the variance of DS was calculated based on the assumption of their null hypothesis $F_1=F_2$, and the test could become less powerful (due to bias in the estimation of the variance) if that null hypothesis does not hold.

\begin{figure}[H]
	\centering
	\includegraphics[scale = 0.8]{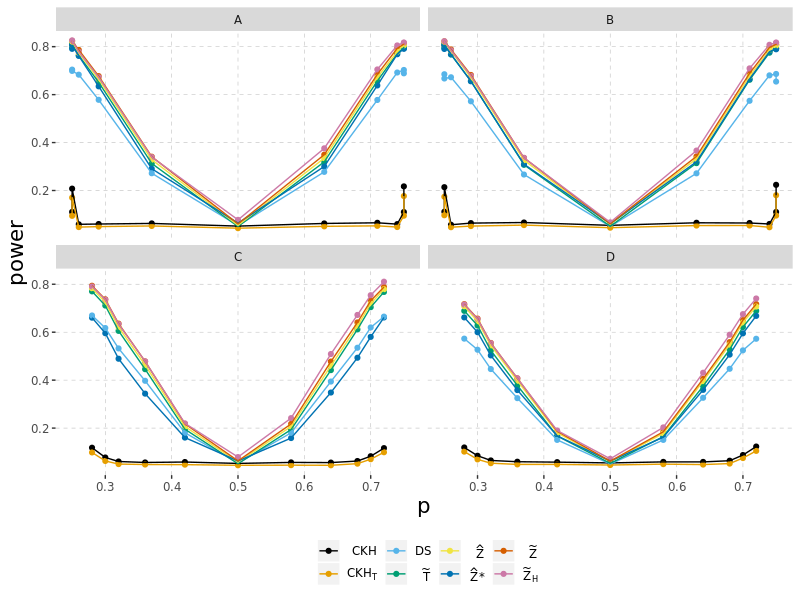}
	{ \caption{\label{fig:powerICS} Plot of achieved power by ratio of theoretical variances of the two groups for all 8 tests.  Panels A and B show multivariate Gaussian cases, while panels C and D shows multivariate Cauchy cases.  This plot serves as a visualization of Tables \ref{table7} and \ref{table8}, as well as Tables S4 and S5 in the supplementary material.  CKH and ${\rm CKH}_T$ are insensitive on detecting the deviation of $p$ from 0.5 under this setting (as mentioned in Sections \ref{sec:motivation} and \ref{subsec:poweranalysis}) when cluster size is informative.}}
\end{figure}

\subsection{Coverage Probability}\label{subsec:covrateanalysis}
It was mentioned in Section \ref{subsec:testintro} that confidence intervals can be obtained from inverting our tests.  In this section we investigate how accurate these interval estimates are.  We will compare our five tests together with CKH in terms of coverage rate of the true parameters.  Note that an interval estimation is not available from DS by its nature.  Tables \ref{table11} and \ref{table12}, as well as Tables S6 and S7 in the supplementary material give results where we consider different distributions and different combinations of $(\rho_1,\rho_2,\rho_{12})$.  These settings are similar to Tables \ref{table7} and \ref{table8}, as well as Tables S4 and S5 in the supplementary material.  In these tables, an additional column named $\text{CKH}_0$ is added to show the coverage rate of CKH for their intended treatment effect $p_0$ under ignorable cluster size.  It can be seen that all five of our tests $\widetilde{Z}$, $\widetilde{T}$, $\widehat{Z}^*$, $\widehat{Z}$, and $\widetilde{Z}_H$ have very similar coverage rates from around 0.92 to 0.94, with $\widetilde{T}$ performing slightly better than the others, which is expected and further confirms the excellence of our small-sample approximation.  Also, we can still see that it is more likely for $\widetilde{Z}_H$ to fail when $p$ is large.  Moreover, CKH has a poor coverage rate for $p$, which is in accordance with what we have seen before.  In contrast, its coverage rate gets noticeably better for $p_0$, which confirms that it is indeed a test for $p_0$ instead of $p$ as mentioned earlier.  
\begin{table}[H]
	\begin{center}
		\begin{tabular}{||c|c|c|c|c|c|c|c|c|c|c|c||} 
			\hline
			\multicolumn{12}{||c||}{$(n_1,n_2,n_c) = (20,10,10)$}\\
			\multicolumn{12}{||c||}{$(\sigma_1^2, \sigma_2^2, \rho_1, \rho_2, \rho_{12}) = (1,1,0.1,0.1,0.9)$}\\
			\hline
			$c_1$ & $c_2$ & $\mu_d$ & $p$ & $p_0$ & $\widetilde{Z}$ & $\widetilde{T}$ & $\widehat{Z}^*$ & $\widehat{Z}$ & $\widetilde{Z}_{H}$ & CKH & $\text{CKH}_0$\\[0.2ex]
			\hline
			2 & 2 & 0 & 0.50 & 0.50 & 93.92 & 95.01 & 93.60 & 94.54 & 92.24 & 94.85 & 94.85\\
			\hline
			2 & 3 & -1 & 0.63 & 0.53 & 93.55 & 94.50 & 93.83 & 93.99 & 92.19 & 79.49 & 94.19\\
			\hline
			3 & 2 & 1 & 0.37 & 0.47 & 93.73 & 94.94 & 93.56 & 94.35 & 93.90 & 79.55 & 94.12\\
			\hline
			2 & 4 & -2 & 0.71 & 0.52 & 93.19 & 94.11 & 93.73 & 93.59 & 90.91* & 51.74 & 93.54\\
			\hline
			4 & 2 & 2 & 0.29 & 0.48 & 93.73 & 94.50 & 93.67 & 94.04 & 93.73 & 52.31 & 93.89\\
			\hline
			2 & 5 & -3 & 0.74 & 0.48 & 92.82 & 93.67 & 93.29 & 93.09 & 89.64* & 31.88 & 94.52\\
			\hline
			5 & 2 & 3 & 0.26 & 0.52 & 93.13 & 94.05 & 93.47 & 93.29 & 93.08 & 31.32 & 94.72\\
			\hline
			2 & 6 & -4 & 0.75 & 0.44 & 91.71 & 92.80 & 92.27 & 91.98 & 88.60* & 18.18 & 94.04\\
			\hline
			6 & 2 & 4 & 0.25 & 0.56 & 92.64 & 93.57 & 93.06 & 92.75 & 92.60 & 17.74 & 94.40\\
			\hline
			2 & 7 & -5 & 0.75 & 0.40 & 92.06 & 93.31 & 92.77 & 92.25 & 88.90* & 9.69 & 94.11\\
			\hline
			7 & 2 & 5 & 0.25 & 0.60 & 92.46 & 93.56 & 92.61 & 92.74 & 92.68 & 9.05 & 94.37\\
			\hline
		\end{tabular}\\
		{ {\caption{\small\label{table11} Coverage rate (\%) under informative cluster size for the treatment effect $p$ when both groups follow multivariate Gaussian distributions. $n_1$, $n_2$ are the number of incomplete clusters of group 1, and 2, respectively.  $n_c$ is the number of complete clusters.  The parameters $\sigma_1^2$, $\sigma_2^2$, $\rho_1$, $\rho_2$ and $\rho_{12}$ are directly associated with the variance-covariance matrix of the distributions, as mentioned in Section \ref{subsec:simsetup}.  $p$ and $p_0$ are the theoretical values of the WMW treatment effect under informative ($p$) and ignorable cluster size ($p_0$), respectively.  The parameters $c_1$ and $c_2$ control the informativeness of the cluster size.  The case $c_1=c_2$ is noninformative (igorable) cluster size.  $\mu_d$ is the difference in theoretical means between the two groups.  The methods $\widetilde{Z}$, $\widetilde{T}$, $\widehat{Z}^*$, $\widehat{Z}$ and $\widetilde{Z}_{H}$ are the proposed method and its variants.  DS is the test proposed in \cite{DS2005} whereas CKH is the test by \citet{CKH2021}.  ${\rm CKH}_0$ shows the coverage rate of the test by \cite{CKH2021} for covering $p_0$, instead of $p$.  The superscript $*$ on the numbers in the body of the table indicates at least 1 (out of 10,000 simulations) result is not available due to negative variance estimation.}}}
	\end{center}
\end{table}

\newpage
\begin{table}[H]
	\begin{center}
		\begin{tabular}{||c|c|c|c|c|c|c|c|c|c|c|c||} 
			\hline
			\multicolumn{12}{||c||}{$(n_1,n_2,n_c) = (20,10,10)$}\\
			\multicolumn{12}{||c||}{$(\sigma_1^2, \sigma_2^2, \rho_1, \rho_2, \rho_{12}) = (1,1,0.9,0.9,0.1)$}\\
			\hline
			$c_1$ & $c_2$ & $\mu_d$ & $p$ & $p_0$ & $\widetilde{Z}$ & $\widetilde{T}$ & $\widehat{Z}^*$ & $\widehat{Z}$ & $\widetilde{Z}_{H}$ & CKH & $\text{CKH}_0$ \\[0.2ex]
			\hline
			2 & 2 & 0 & 0.50 & 0.50 & 93.82 & 94.91 & 93.81 & 94.15 & 93.28 & 94.53 & 94.53\\
			\hline
			2 & 3 & -1 & 0.63 & 0.53 & 93.48 & 94.41 & 93.38 & 93.84 & 92.01 & 82.36 & 94.23\\
			\hline
			3 & 2 & 1 & 0.37 & 0.47 & 93.46 & 94.49 & 93.68 & 93.83 & 93.45 & 82.43 & 93.89\\
			\hline
			2 & 4 & -2 & 0.71 & 0.52 & 92.95 & 93.85 & 93.40 & 93.29 & 90.71* & 56.18 & 93.76\\
			\hline
			4 & 2 & 2 & 0.29 & 0.48 & 93.39 & 94.16 & 93.48 & 93.66 & 93.45 & 55.73 & 94.27\\
			\hline
			2 & 5 & -3 & 0.74 & 0.48 & 92.76 & 93.70 & 92.96 & 93.02 & 89.47* & 33.40 & 94.45\\
			\hline
			5 & 2 & 3 & 0.26 & 0.52 & 92.80 & 93.79 & 93.27 & 93.06 & 92.83 & 32.64 & 94.80\\
			\hline
			2 & 6 & -4 & 0.75 & 0.44 & 91.82 & 92.87 & 92.24 & 92.01 & 88.56* & 18.25 & 93.85\\
			\hline
			6 & 2 & 4 & 0.25 & 0.56 & 92.64 & 93.56 & 93.03 & 92.75 & 92.63 & 17.89 & 94.43\\
			\hline
			2 & 7 & -5 & 0.75 & 0.40 & 92.05 & 93.41 & 92.85 & 92.34 & 88.99* & 9.69 & 93.90\\
			\hline
			7 & 2 & 5 & 0.25 & 0.60 & 92.49 & 93.60 & 92.71 & 92.77 & 92.77 & 8.95 & 94.23\\
			\hline
		\end{tabular}\\
		{ {\caption{\small\label{table12} Coverage rate (\%) under informative cluster size for the treatment effect $p$ when both groups follow multivariate Gaussian distributions. $n_1$, $n_2$ are the number of incomplete clusters of group 1, and 2, respectively.  $n_c$ is the number of complete clusters.  The parameters $\sigma_1^2$, $\sigma_2^2$, $\rho_1$, $\rho_2$ and $\rho_{12}$ are directly associated with the variance-covariance matrix of the distributions, as mentioned in Section \ref{subsec:simsetup}.  $p$ and $p_0$ are the theoretical values of the WMW treatment effect under informative ($p$) and ignorable cluster size ($p_0$), respectively.  The parameters $c_1$ and $c_2$ control the informativeness of the cluster size.  The case $c_1=c_2$ is noninformative (igorable) cluster size.  $\mu_d$ is the difference in theoretical means between the two groups.  The methods $\widetilde{Z}$, $\widetilde{T}$, $\widehat{Z}^*$, $\widehat{Z}$ and $\widetilde{Z}_{H}$ are the proposed method and its variants.  DS is the test proposed in \cite{DS2005} whereas CKH is the test by \citet{CKH2021}.  ${\rm CKH}_0$ shows the coverage rate of the test by \cite{CKH2021} for covering $p_0$, instead of $p$.  The superscript $*$ on the numbers in the body of the table indicates at least 1 (out of 10,000 simulations) result is not available due to negative variance estimation.  In particular, simulation results from this table uses a slightly different scale matrix compared to Table \ref{table11}.}}}
	\end{center}
\end{table}

Here again, we may recall Figure \ref{fig:introcomparison}. It can be clearly seen in the plot that the curve CKH for $p_0$ is constantly close to the dotted blue line of 0.95, whereas the coverage rate of CKH for $p$ drops significantly as the absolute difference between $p$ and $p_0$ gets large.  The visualization agrees with our previous findings.

\subsection{Summary of the Simulation Results}
From the simulation results, it can be seen that CKH performs poorly under informative cluster size, and DS fails to perform well when the null hypothesis of $p = \frac{1}{2}$ holds with the violation of $F_1=F_2$.  Even though on average $\widehat{Z}^*$ performs better than expected, just a single result of it could often be unreliable since it is only based on one resample.  The tests $\widehat{Z}$ and $\widetilde{Z}_{H}$ have some potentials, but it cannot be guaranteed that their estimated variances would always be positive, especially for $\widetilde{Z}_{H}$ when true $p$ is large.  This can lead to occasional unavailability of results.  Also, the achieved type I error rate is usually inflated for $\widetilde{Z}_{H}$ under our $H_0$.  Overall, under the assumption of informative cluster size, our test $\widetilde{Z}$ performs the most favorably.  The small-sample approximation $\widetilde{T}$ should be used in general, and especially when the estimated degrees of freedom is less than 30, or when the cluster sizes or number of clusters are small.  

 Our tests still provides competitive results in the case where cluster size is not informative. The reason is that if cluster size does not have an effect, then the reweighting process of WCR uses roughly uniform weight and, thus, our calculations and derivations still remain valid.  This provides support for using our methods even when we are uncertain and would like to know if informative cluster size is in presence.  More specifically, if our test result is very different from that of CKH, it counts as strong indication of presence of informative cluster size.  In addition, our test will work under the most restrictive settings with no clustering effect, because this is essentially the special case of either $m_i^{(1)} = m_i = 1$ or $m_i^{(2)} = m_i = 1$ for all $i=1,\cdots,n$ and $n_c = 0$.



\section{Real Data Analysis}\label{sec:realdata}
In this section, we analyze two real datasets: one with non-informative cluster size and the other one with informative cluster size.

\subsection{Alcohol Use Data - Ignorable Cluster Size}
This longitudinal data, originally from \citet{CSC1997}, is about alcohol use among  teens.  The original data contains observations from $363$ adolescents, but here we only included $82$ subjects who were originally of age $14$ at the beginning of the study (see also \cite{SW2003}).  This data is clustered in a sense that each subject can be  considered as a cluster where observations are taken over $3$ years.  The available data consists of $246$ observations, with exactly $3$ observations taken on each subject.   Here, the cluster sizes are balanced and, thus, are uninformative.  One particular risk factor of interest is COA, which is a binary variable that indicates whether or not at least one of the subject's parents is alcoholic.  The outcome variable is the amount of alcohol consumed in terms of frequency with which they (1) drank beer or wine, (2) drank hard liquor, (3) had five or more drinks in a row, and (4) got drunk using an 8-point scale (ranging from 0 = “not at all” to 7 = “every day”).  The outcome variable used in this manuscript is a score computed by taking the square root of the sum of all responses for each subject (see \cite{SW2003} for more details).  Figure \ref{figure4} shows histograms of the outcome variable of the two groups.  It can be seen that the data is clearly skewed.  The second group has larger density to the right compared to the first group, indicating that the adolescents with at least one alcoholic parent tend to use more alcohol.
	
\begin{figure}[H]
	\centering
	\includegraphics[scale = 0.8]{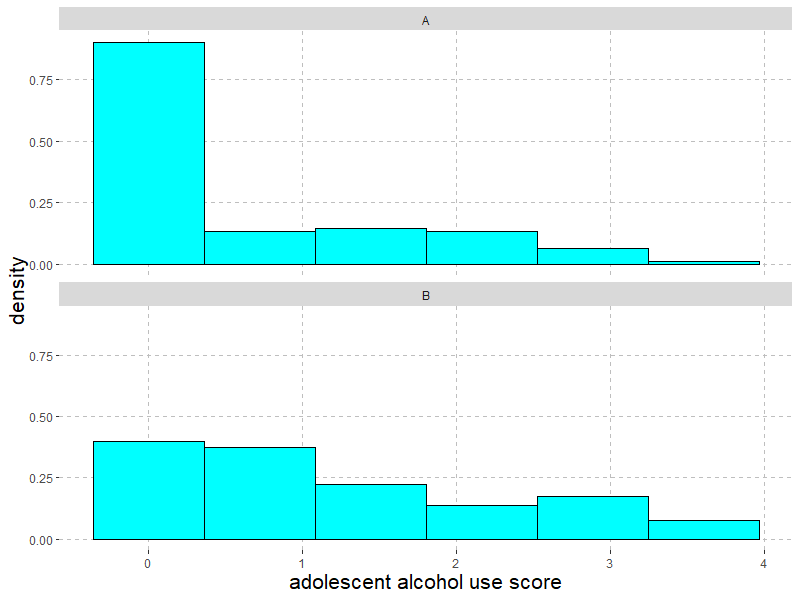}
	\caption{\label{figure4} Histogram of the adolescent alcohol use data: panel A - distribution of alcohol use in group 1 (with no alcoholic parents); panel B - distribution of alcohol use in group 2 (with at least one alcoholic parents).}
\end{figure}

All the methods evaluated in Section \ref{sec:sim} are applied and the results are summarized in Table \ref{table15}.  It can be seen that the estimated treatment effects are very similar (around 0.68) for all methods, which is expected when data has noninformative cluster size.  The estimation of the treatment effect is larger than 0.5, meaning that there is a high probability that the alcohol use of group 1 (adolescents with no alcoholic parent) is less than the alcohol use of group 2 (adolescents with at least one alcoholic parent), which is consistent with Figure \ref{figure4} and one would naturally expect.  In particular, the estimated treatment effects from our $\widetilde{p}$ and CKH method are exactly the same, since there is no difference between weighting on each observation and on each cluster in this example.  Also, $p-$values of all tests are very small, meaning that the data provide sufficient evidence that COA has an impact on adolescent alcohol use according to all tests. Especially, the small-sample approximations have larger $p-$value than their large-sample counterparts.  The largest $p-$value from $\widehat{Z}^*$ is possibly because of its comparatively large variance estimate.  The confidence interval estimations are also very similar across all tests.

\begin{table}[H]
	\begin{center}
		\begin{tabular}{||c|c|c|c|c|c|c|c|c||} 
			\hline
			& $\widetilde{Z}$ & $\widetilde{T}$ & $\widehat{Z}^*$ & $\widehat{Z}$ & $\widetilde{Z}_H$ & DS & CKH & $\text{CKH}_T$\\[0.2ex]
			\hline
			estimated $p$ & 0.6823 & 0.6823 & 0.6685 & 0.6814 & 0.6823 & - & 0.6823 & 0.6823\\
			\hline
			$p-$value & $5.4e^{-5}$ & $1.3e^{-4}$ & $3.0e^{-3}$ & $6.1e^{-5}$ & $5.7e^{-5}$ & $2.4e^{-4}$ & $5.4e^{-5}$ & $1.2e^{-4}$\\
			\hline
			95\% LB & 0.5938 & 0.5924 & 0.5572 & 0.5927 & 0.5935 & - & 0.5939 & 0.5925\\
			\hline
			95\% UB & 0.7709 & 0.7723 & 0.7797 & 0.7701 & 0.7712 & - & 0.7708 & 0.7722\\
			\hline
		\end{tabular}\\
		{ \caption{\label{table15} Estimated treatment effects, $p-$values, and interval estimates performed on the adolescent alcohol use data.  The first 5 columns show results of our tests.  DS is the test proposed in \cite{DS2005}; CKH and ${\rm CKH}_T$ are the test and its small-sample approximation as mentioned in \cite{CKH2021}.  Point and interval estimations are not viable for \cite{DS2005}.}}
	\end{center}
\end{table}

\subsection{Periodontal Data - Informative Cluster Size}
The data set analyzed in this section is a subset taken from a dental study of attachment loss from \cite{BSKO1997}.  The data was collected by the Piedmont 65+ longitudinal dental study initiated by the School of Dentistry of the University of North Carolina over a 5-year study on 1,000 subjects.  The partial data that we have contains 807 subjects.  After further investigation, we selected 27 subjects who had been smoking and had no missing data on record so that a missing tooth is almost certainly due to health reasons and not for other reasons.  The outcome variable is the attachment loss, which is a score measured by the sum of attachment loss (in millimeters) of both mid-buccal and mesio-buccal sites of a particular tooth. 
 High attachment loss score indicates poor overall oral health condition of the subject.  In this analysis, we compare the attachment loss at baseline (group 1) vs. at 18-month follow-up (group 2) with no intervention to see whether or not smoking has an effect on the oral health of the subjects.  Each subject is considered a cluster, where observations are taken from all $32$ teeth of each individual.   Note that there is dependence among teeth on a same subject, and also data from the same tooth at baseline and follow-up are related.  No data would be recorded on a tooth if the subject is missing that particular tooth.  The partial data of interest consists of observations obtained from $470$ teeth of $27$ subjects at two time points.  Figure \ref{figure5} shows histograms of the outcome variable of the two groups.  Again, it can be seen that the data is skewed, so nonparametric method is the appropriate method of analysis. The histograms show the attachment loss of all available teeth, pulling all subjects.  Not much can be said just by looking at the plot since there is no noticeable difference between the two plots, and it is difficult to tell which distribution tends to provide larger observations.

\begin{figure}[H]
	\centering
	\includegraphics[scale = 0.8]{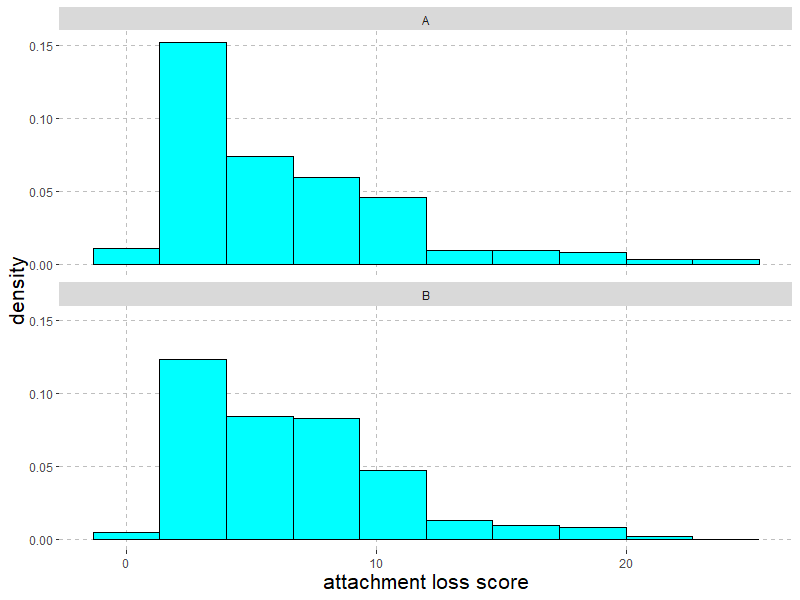}
	\caption{\label{figure5} Histogram of the periodontal data: panel A - distribution of attachment loss in group 1 (baseline); panel B - distribution of alcohol use in group 2 (18 month follow-up).}
\end{figure}

 Table \ref{table16} contains the results of the analysis.  It can be seen that there is a difference in the estimation of the treatment effect $p$ between our methods and that of CKH, even though almost all $p-$values are small.  The fact that this estimated treatment effect is larger than 0.5 shows that the attachment loss from group 1 tend to be smaller than group 2, which is clinically sensible as a lot of research have shown that smoking deteriorates both oral and overall health of an individual.  More precisely, our methods estimate $p$ to be around 0.57, whereas CKH estimates $p$ to be around 0.55.  This means that the treatment effect tends to be more obvious if we take into consideration the informativeness of cluster sizes.  On the other hand, we see that $\widehat{Z}^*$ gives a quite different result here with a very large $p-$value and a very wide confidence interval.  This is likely due to large variance estimate resulting from using only a single resample.  Overall, the information in cluster sizes in this particular dataset does not make a very big difference in the estimation of treatment effect $p$.

\begin{table}[H]
	\begin{center}
		\begin{tabular}{||c|c|c|c|c|c|c|c|c||} 
			\hline
			& $\widetilde{Z}$ & $\widetilde{T}$ & $\widehat{Z}^*$ & $\widehat{Z}$ & $\widetilde{Z}_H$ & DS & CKH & $\text{CKH}_T$\\[0.2ex]
			\hline
			estimated $p$ & 0.5721 & 0.5721 & 0.4972 & 0.5713 & 0.5721 & - & 0.5546 & 0.5546\\
			\hline
			$p-$value & $3.8e^{-4}$ & $1.5e^{-3}$ & $0.98$ & $2.8e^{-2}$ & $4.3e^{-2}$ & $2.8e^{-3}$ & $5.9e^{-4}$ & $2.0e^{-3}$\\
			\hline
			95\% LB & 0.5323 & 0.5303 & 0.2723 & 0.5076 & 0.5022 & - & 0.5234 & 0.5219\\
			\hline
			95\% UB & 0.6119 & 0.6139 & 0.7220 & 0.6351 & 0.6421 & - & 0.5857 & 0.5873\\
			\hline
		\end{tabular}\\
		{ \caption{\label{table16} Estimated treatment effects, $p-$values, and interval estimates performed on the the partial Piedmont 65+ periodontal data.  The first 5 columns show results of our tests.  DS is the test proposed in \cite{DS2005}; CKH and ${\rm CKH}_T$ are the test and its small-sample approximation as mentioned in \cite{CKH2021}.  Point and interval estimations are not viable for \cite{DS2005}.}}
	\end{center}
\end{table}



\section{Summary and Conclusion}\label{sec:conclusion}
In this paper, we investigated Wilcoxon-Mann-Whitney effect (WMW) in the cluster data framework with informative (nonignorable) cluster size.  Cluster size is informative if the distribution of the outcome depends on cluster size.  The parameter targeted conceptually differ under informative and noninformative cluster.  Methods designed under ignorable cluster size perform poorly under informative cluster size.  

We proposed unbiased estimator for the WMW effect and investigated asymptotic properties.  The construction of our estimator is based on within-cluster resampling. We proved the consistency and asymptotic normality the estimator.  Our theoretical approach relies on the idea of Hájek’s projection to obtain asymptotically equivalent version of the estimator.  The asymptotic distribution and its variance are derived from the projection.  Ratio-consistent estimator for the asymptotic variance is also given.  An alternative variance estimator can be constructed from a conditional variance argument.  However, this estimator has the undesirable feature of taking negative values, albeit rarely. 

To the best of our knowledge, this is the first work considering the WMW effect under informative cluster size.  There are a few works that have studied nonparametric tests under informative cluster size but for testing the hypothesis of equality of distributions.   Our numerical results have demonstrated that the proposed estimator and test based on it performs considerably better than estimators proposed under ignorable cluster size.   

Another possible approach for constructing an estimator is taking the average of WMW estimators calculated from all possible resamples as in \eqref{eq:p_hat}.  This estimator coincides with our proposed estimator if there are no complete clusters.  The general case is more involved.  This paper considers only a two-group problem.  The extension to the multiple group situation is in principle similar but the formulation of the effect size and its estimators need some care.  We defer these extensions to a future manuscript.  



\section*{Acknowledgement}
The authors are very grateful for Dr. Jim Beck and Dr. Kevin Moss for their willingness and time providing the Piedmont 65+ periodontal data.  The research of S. W. Harrar was carried out while visiting the Institute of Biometry and Clinical Epidemiology at Charit\'e University of Medicine, Berlin, Germany as a guest scientist.  He is grateful to the institute for the wonderful research environment and hospitality. 



\newpage
\bibliographystyle{abbrvnat} 
\bibliography{References.bib} 

\begin{thebibliography}{18}
\providecommand{\natexlab}[1]{#1}
\providecommand{\url}[1]{\texttt{#1}}
\expandafter\ifx\csname urlstyle\endcsname\relax
  \providecommand{\doi}[1]{doi: #1}\else
  \providecommand{\doi}{doi: \begingroup \urlstyle{rm}\Url}\fi

\bibitem[Beck et~al.(1997)Beck, Sharp, Koch, and Offenbacher]{BSKO1997}
J.~D. Beck, T.~Sharp, G.~G. Koch, and S.~Offenbacher.
\newblock A study of attachment loss patterns in survivor teeth at 18 months,
  36 months and 5 years in community-dwelling older adults.
\newblock \emph{Journal of periodontal research}, 32\penalty0 (6):\penalty0
  497--505, 1997.

\bibitem[Brunner and Munzel(2000)]{BM2000}
E.~Brunner and U.~Munzel.
\newblock The nonparametric behrens-fisher problem: Asymptotic theory and a
  small-sample approximation.
\newblock \emph{Biometrical Journal}, 42\penalty0 (1):\penalty0 17--25, 2000.

\bibitem[Brunner et~al.(1997)Brunner, Dette, and Munk]{BDM1997}
E.~Brunner, H.~Dette, and A.~Munk.
\newblock Box-type approximations in nonparametric factorial designs.
\newblock \emph{Journal of the American Statistical Association}, 92\penalty0
  (440):\penalty0 1494--1502, 1997.

\bibitem[Brunner et~al.(2017)Brunner, Konietschke, Pauly, and Puri]{BKPP2017}
E.~Brunner, F.~Konietschke, M.~Pauly, and M.~L. Puri.
\newblock Rank-based procedures in factorial designs: Hypotheses about
  non-parametric treatment effects.
\newblock \emph{Journal of the Royal Statistical Society: Series B (Statistical
  Methodology)}, 79\penalty0 (5):\penalty0 1463--1485, 2017.

\bibitem[Cui et~al.(2021)Cui, Konietschke, and Harrar]{CKH2021}
Y.~Cui, F.~Konietschke, and S.~W. Harrar.
\newblock The nonparametric behrens–fisher problem in partially complete
  clustered data.
\newblock \emph{Biometrical Journal}, 63\penalty0 (1):\penalty0 148--167, 2021.
\newblock \doi{https://doi.org/10.1002/bimj.201900310}.
\newblock URL
  \url{https://onlinelibrary.wiley.com/doi/abs/10.1002/bimj.201900310}.

\bibitem[Curran et~al.(1997)Curran, Stice, and Chassin]{CSC1997}
P.~J. Curran, E.~Stice, and L.~Chassin.
\newblock The relation between adolescent and peer alcohol use: A longitudinal
  random coefficients model.
\newblock \emph{Journal of Consulting and Clinical Psychology}, 65:\penalty0
  130--140, 1997.

\bibitem[Datta and Satten(2005)]{DS2005}
S.~Datta and G.~A. Satten.
\newblock Rank-sum tests for clustered data.
\newblock \emph{Journal of the American Statistical Association}, 100\penalty0
  (471):\penalty0 908--915, 2005.

\bibitem[Datta and Satten(2008)]{DS2008}
S.~Datta and G.~A. Satten.
\newblock A signed-rank test for clustered data.
\newblock \emph{Biometrics}, 64\penalty0 (2):\penalty0 501--507, 2008.

\bibitem[Dutta and Datta(2016)]{DD2016}
S.~Dutta and S.~Datta.
\newblock A rank-sum test for clustered data when the number of subjects in a
  group within a cluster is informative.
\newblock \emph{Biometrics}, 72\penalty0 (2):\penalty0 432--440, 2016.

\bibitem[Fligner and Policello(1981)]{FP1981}
M.~A. Fligner and G.~E. Policello.
\newblock Robust rank procedures for the behrens-fisher problem.
\newblock \emph{Journal of the American Statistical Association}, 76\penalty0
  (373):\penalty0 162--168, 1981.

\bibitem[Hoffman et~al.(2001)Hoffman, Sen, and Weinberg]{HSW2001}
E.~B. Hoffman, P.~K. Sen, and C.~R. Weinberg.
\newblock Within-cluster resampling.
\newblock \emph{Biometrika}, 88\penalty0 (4):\penalty0 1121--1134, 2001.

\bibitem[Larocque et~al.(2010)Larocque, Haataja, Nevalainen, and Oja]{LHNO2010}
D.~Larocque, R.~Haataja, J.~Nevalainen, and H.~Oja.
\newblock Two sample tests for the nonparametric behrens–fisher problem with
  clustered data.
\newblock \emph{Journal of Nonparametric Statistics}, 22\penalty0 (6):\penalty0
  755--771, 2010.
\newblock \doi{10.1080/10485250903469728}.
\newblock URL \url{https://doi.org/10.1080/10485250903469728}.

\bibitem[Mann and Whitney(1947)]{MW1947}
H.~B. Mann and D.~R. Whitney.
\newblock On a test of whether one of two random variables is stochastically
  larger than the other.
\newblock \emph{The Annuals of Mathematical Statistics}, 18\penalty0
  (1):\penalty0 50--60, 1947.
\newblock \doi{https://doi.org/10.1214/aoms/1177730491}.

\bibitem[Rosner et~al.(2004)Rosner, Glynn, and Lee]{RGL2004}
B.~Rosner, R.~Glynn, and M.-L. Lee.
\newblock Incorporation of clustering effects for the wilcoxon rank sum test: A
  large-sample approach.
\newblock \emph{Biometrics}, 59:\penalty0 1089--1098, 01 2004.
\newblock \doi{10.1111/j.0006-341X.2003.00125.x}.

\bibitem[Roy et~al.(2019)Roy, Harrar, and Konietschke]{RHK2019}
A.~Roy, S.~W. Harrar, and F.~Konietschke.
\newblock The nonparametric behrens-fisher problem with dependent replicates.
\newblock \emph{Statistics in Medicine}, 38\penalty0 (25):\penalty0 4939--4962,
  2019.
\newblock \doi{https://doi.org/10.1002/sim.8343}.
\newblock URL \url{https://onlinelibrary.wiley.com/doi/abs/10.1002/sim.8343}.

\bibitem[Singer and Willett(2003)]{SW2003}
J.~D. Singer and J.~B. Willett.
\newblock \emph{Applied Longitudinal Data Analysis: Modeling Change and Event
  Occurrence}.
\newblock Oxford University Press, 1 edition, 2003.

\bibitem[Wilcoxon(1945)]{W1945}
F.~Wilcoxon.
\newblock Individual comparisons by ranking methods.
\newblock \emph{Biometrics Bulletin}, 1\penalty0 (6):\penalty0 80--83, 1945.

\bibitem[Williamson et~al.(2003)Williamson, Datta, and Satten]{WDS2003}
J.~M. Williamson, S.~Datta, and G.~A. Satten.
\newblock Marginal analyses of clustered data when cluster size is informative.
\newblock \emph{Biometrics}, 59\penalty0 (1):\penalty0 36--42, 2003.

\end{thebibliography}


\end{document}


\allowdisplaybreaks
\author{Changrui Liu and Solomon W.  Harrar\\Dr. Bing Zhang Department of Statistics\\
University of Kentucky, Lexington, KY 40506, USA} 

\title{Supplementary Material\\ {Wilcoxon-Mann-Whitney Effects for Clustered Data: Informative Cluster Size}} 

\maketitle


\section{Proofs}\label{app:proofs}

This section contains the proofs of the results presented in the main manuscript.  Some additional results needed for the proofs are also stated as propositions in this supplementary material and  their proofs are given.  
\begin{proof}[Proof of Proposition \ref{prop:ptilde}]
    \begin{itemize}
        \item[(i)]  Unbiasedness:  In view of \eqref{eq:p_tilde}, it suffices to show that
    \begin{align*}
    \bigg(p-\frac{1}{2}\bigg)\sum_{i=1}^{n}\sum_{j\ne i}\alpha_{i}(1-\alpha_{j}) = \frac{n-1}{2}\sum\limits_{i=1}^{n} \alpha_i - \sum\limits_{i=1}^{n}\sum\limits_{j\ne i}\sum\limits_{k=1}^{m_j} \frac{\alpha_i}{m_j} E[\widehat{F}_{2i}(X_{jk})].
\end{align*}
However, the above step is trivial noting that $E[\widehat{F}_{2i}(X_{jk})] = 1-p$ for $k = 1,\cdots,m_j^{(1)}$, and $E[\widehat{F}_{2i}(X_{jk})] = \frac{1}{2}$ for $k = m_j^{(1)},\cdots,m_j$, regardless of $i$ and $j$.

\item[(ii)]  Consistency: 
   Since $\widetilde{p}$ is unbiased, to prove consistency it suffices to show that 
    $$Var(\widetilde{p})\rightarrow 0 \text{ as } n\rightarrow\infty\text{.}$$
 With the $R_1$ and $R_2$ notations defined  in Section \ref{sec:estimation}, $\widetilde{p}$ can be re-expressed as
\begin{equation}
    \widetilde{p} = \frac{1}{E(m_1^*m_2^*)}\sum_{i\in R_1}\sum_{j\in R_2, j\ne i} (1-\alpha_i)\alpha_j \int \widehat{F}_{1i} d\widehat{F}_{2j}\label{ptildealtform},
\end{equation}
where $\widehat{F}_{1i}$ denotes the normalized empirical distribution function for group 1 within cluster $i$, similar to its counterpart $\widehat{F}_{2j}$, and 
\begin{align*}
    \int \widehat{F}_{1i}d\widehat{F}_{2j}
    &= \frac{1}{m_i^{(1)}m_j^{(2)}}\sum_{k=1}^{m_i^{(1)}}\sum_{h=m_j^{(1)}+1}^{m_j}\bigg[\frac{1}{2}\bigg(I(X_{ik}<X_{jh}) + I(X_{ik}\le X_{jh})\bigg)\bigg].
\end{align*}

Starting from expression \eqref{ptildealtform}, we have that 
{\begin{align*}
    Var(\widetilde{p}) &= Var\bigg[\frac{1}{E(m_1^*m_2^*)}\sum_{i\in R_1}\sum_{j\ne i, j\in R_2} (1-\alpha_i)\alpha_j\int \widehat{F}_{1i} d\widehat{F}_{2j}\bigg]\\
    &= [E(m_1^*m_2^*)]^{-2} \sum_{i_1\in R_1}\sum_{j_1\ne i_1, j_1\in R_2}\sum_{i_2\in R_1}\sum_{j_2\ne i_2, j_2\in R_2}\\
    &\hspace{1cm}(1-\alpha_{i_1})\alpha_{j_1}(1-\alpha_{i_2})\alpha_{j_2}Cov\bigg(\int \widehat{F}_{1i_1} d\widehat{F}_{2j_1}, \int \widehat{F}_{1i_2} d\widehat{F}_{2j_2}\bigg).
\end{align*}}
Note that 
\begin{align*}
    E(m_1^*m_2^*) &= (n-1)\sum \alpha_i + \sum \alpha_i^2 - \bigg(\sum \alpha_i\bigg)^2.
\end{align*}
By Condition \ref{cond:numclus}, $E(m_1^*m_2^*)$ is of order $n^2$.  It now suffices to show that the order of the four summation covariance term is less than $n^4$.  To this end, there are four scenarios to consider:
\begin{enumerate}
    \item $i_1\ne i_2$, $j_1\ne j_2$.
   In this case, $Cov(\int \widehat{F}_{1i_1} d\widehat{F}_{2j_1}, \int \widehat{F}_{1i_2} d\widehat{F}_{2j_2})=0$ by independence.
    
     \item $i_1=i_2=i$, $j_1\ne j_2$.
    We have that
    \begin{align*}
        &Cov\bigg(\int \widehat{F}_{1i_1} d\widehat{F}_{2j_1}, \int \widehat{F}_{1i_2} d\widehat{F}_{2j_2}\bigg)\\
        &= E\bigg\{\frac{1}{m_{i}^{(1)}m_{j_1}^{(2)}}\sum_{k_1 = 1}^{m_{i}^{(1)}}\sum_{h_1 = m_{j_1}^{(1)}+1}^{m_{j_1}}\bigg[\frac{1}{2}\bigg(I(X_{ik_1}<X_{j_1h_1}) + I(X_{ik_1}\le X_{j_1h_1})\bigg)\bigg]\times\\
        &\hspace{1cm} \frac{1}{m_{i}^{(1)}m_{j_2}^{(2)}}\sum_{k_2 = 1}^{m_i^{(1)}}\sum_{h_2 = m_{j_2}^{(1)}+1}^{m_{j_2}}\bigg[\frac{1}{2}\bigg(I(X_{ik_2}<X_{j_2h_1}) + I(X_{ik_2}\le X_{j_2h_2})\bigg)\bigg]\bigg\} - \\
        &\hspace{0.5cm}E\bigg\{\frac{1}{m_{i}^{(1)}m_{j_1}^{(2)}}\sum_{k_1 = 1}^{m_{i}^{(1)}}\sum_{h_1 = m_{j_1}^{(1)}+1}^{m_{j_1}}\bigg[\frac{1}{2}\bigg(I(X_{ik_1}<X_{j_1h_1}) + I(X_{ik_1}\le X_{j_1h_1})\bigg)\bigg]\bigg\}\times\\
        &\hspace{1cm} E\bigg\{\frac{1}{m_{i}^{(1)}m_{j_2}^{(2)}}\sum_{k_2 = 1}^{m_{i}^{(1)}}\sum_{h_1 = m_{j_2}^{(1)}+1}^{m_{j_2}}\bigg[\frac{1}{2}\bigg(I(X_{ik_2}<X_{j_2h_2}) + I(X_{ik_2}\le X_{j_2h_2})\bigg)\bigg]\bigg\}.
    \end{align*}
    Note that given any pair of $i\in R_1,j\in R_2$ where $i\ne j$,
    \begin{align*}
        &E\bigg\{\frac{1}{m_{i}^{(1)}m_{j}^{(2)}}\sum_{k = 1}^{m_{i}^{(1)}}\sum_{h = m_{j}^{(1)}+1}^{m_{j}}\bigg[\frac{1}{2}\bigg(I(X_{ik}<X_{jh}) + I(X_{ik}\le X_{jh})\bigg)\bigg]\bigg\}\\
        &= \frac{1}{m_{i}^{(1)}m_{j}^{(2)}}\sum_{k = 1}^{m_{i}^{(1)}}\sum_{h = m_{j}^{(1)}+1}^{m_{j}}p = p.
    \end{align*}
    On the other hand, 
    \begin{align*}
        &E\bigg\{\frac{1}{m_{i}^{(1)}m_{j_1}^{(2)}}\sum_{k_1 = 1}^{m_{i}^{(1)}}\sum_{h_1 = m_{j_1}^{(1)}+1}^{m_{j_1}}\bigg[\frac{1}{2}\bigg(I(X_{ik_1}<X_{j_1h_1}) + I(X_{ik_1}\le X_{j_1h_1})\bigg)\bigg]\cdot\\
        &\hspace{1cm}\frac{1}{m_{i}^{(1)}m_{j_2}^{(2)}}\sum_{k_2 = 1}^{m_i^{(1)}}\sum_{h_2 = m_{j_2}^{(1)}+1}^{m_{j_2}}\bigg[\frac{1}{2}\bigg(I(X_{ik_2}<X_{j_2h_1}) + I(X_{ik_2}\le X_{j_2h_2})\bigg)\bigg]\bigg\}\\
        &= [(m_{i}^{(1)})^2m_{j_1}^{(2)}m_{j_2}^{(2)}]^{-1}\cdot E\bigg\{E\bigg\{m_i^{(1)}m_{j_1}^{(2)} - \sum_{k = 1}^{m_{i}^{(1)}} \bigg[m_{j_1}^{(2)}\widehat{F}_{2j_1}(X_{ik})\bigg|\bm{X_i}\bigg]\bigg\}\cdot\\
        &\hspace{1cm} E\bigg\{m_i^{(1)}m_{j_2}^{(2)} - \sum_{k = 1}^{m_{i}^{(1)}} \bigg[m_{j_2}^{(2)}\widehat{F}_{2j_2}(X_{ik})\bigg|\bm{X_i}\bigg]\bigg\}\bigg\}\\
        &= 1-2(1-p)+[m_i^{(1)}]^{-2}E\bigg[\bigg(\sum_{k = 1}^{m_{i}^{(1)}}F_{2}(X_{ik})\bigg)^2\bigg].
    \end{align*}
    Also, 
    \begin{align*}
        E\bigg[\bigg(\sum_{k = 1}^{m_{i}^{(1)}}F_{2}(X_{ik})\bigg)^2\bigg]&\le (m_i^{(1)})^2
    \end{align*}
    since each $F_{2}(X_{ik})$ is bounded above by 1 for each $k$.  Therefore, it follows that 
    \begin{align*}
        Cov\bigg(\int \widehat{F}_{1i} d\widehat{F}_{2j_1}, \int \widehat{F}_{1i} d\widehat{F}_{2j_2}\bigg) \le 2p-p^2.
    \end{align*}

    \item $i_1\ne i_2$, $j_1=j_2=j$.
    Along the same lines with the previous case, after some algebra we have
    \begin{align*}
        Cov\bigg(\int \widehat{F}_{1i_1} d\widehat{F}_{2j}, \int \widehat{F}_{1i_2} d\widehat{F}_{2j}\bigg)\le 1-p^2\text{.}
    \end{align*}
    \item $i_1=i_2=i$, $j_1=j_2=j$.  In this case, 
    \begin{align*}
        Var\bigg(\int \widehat{F}_{1i} d\widehat{F}_{2j}\bigg)
        &= E\bigg[\bigg(\int \widehat{F}_{1i} d\widehat{F}_{2j}\bigg)^2\bigg] - E^2\bigg(\int \widehat{F}_{1i} d\widehat{F}_{2j}\bigg)\\
        &\le 1-p^2.
    \end{align*}
\end{enumerate}
From the four cases above, we have 
\begin{align*}
    &\sum_{i_1\in R_1}\sum_{j_1\ne i_1, j_1\in R_2}\sum_{i_2\in R_1}\sum_{j_2\ne i_2, j_2\in R_2}(1-\alpha_{i_1})\alpha_{j_1}(1-\alpha_{i_2})\alpha_{j_2}\\
    &\hspace{1cm} Cov\bigg(\int \widehat{F}_{1i_1} d\widehat{F}_{2j_1}, \int \widehat{F}_{1i_2} d\widehat{F}_{2j_2}\bigg)\\
    &\le \sum_{i\in R_1}\sum_{j_1\ne i, j_1\in R_2}\sum_{j_2\ne i, j_2\ne j_1, j_2\in R_2} (1-\alpha_i)^2\alpha_{j_1}\alpha_{j_2}(2p-p^2) + \\
    &\hspace{1cm} \sum_{j\in R_2}\sum_{i_1\ne j, i_1\in R_1}\sum_{i_2\ne j, i_2\ne i_1, i_2\in R_2} (1-\alpha_{i_1})(1-\alpha_{i_2})\alpha_{j}^2(1-p^2) + \\
    &\hspace{1cm} \sum_{i\in R_1}\sum_{j\ne i, j\in R_2}(1-\alpha_{i})^2\alpha_{j}^2(1-p^2)\le O(n^3).
\end{align*}
Therefore, it follows that $Var(\widetilde{p})\le O(n^{-1})$ and that completes the proof of consistency.
\end{itemize}
\end{proof}

\begin{proof}[Proof of Theorem \ref{thm:ptildenorm}]

    Define
\begin{align*}
	U &= \frac{1}{n} E(U^*|\bm{X,g})\\
	&= \frac{1}{n}E\bigg[\sum_{i=1}^{n}\sum_{j\ne i} (1-g_i^*)g_j^*\cdot\frac{1}{2}\bigg(I(X^*_{i} < X^*_{j}) +  I(X^*_{i} \le X^*_{j})\bigg)\bigg|\bm{X,g}\bigg]\\
	&= \sum_{i=1}^{n}\alpha_i - \frac{1}{n}\sum_{i=1}^{n}\sum_{j\ne i}\sum_{k=1}^{m_j} \frac{\alpha_i}{m_j} \widehat{F}_{2i}(X_{jk}) - \frac{1}{2n}\bigg[\sum_{i=1}^{n}\alpha_i(1-\alpha_i)+ \bigg(\sum_{i=1}^{n}\alpha_i\bigg)^2+\sum_{i=1}^{n}\alpha_i\bigg].
\end{align*}
Since 
$$\frac{\widetilde{p}-p}{\sqrt{Var(\widetilde{p})}} = \frac{U-E(U)}{\sqrt{Var(U)}},$$
we will be done if we show that the latter converges in distribution to $N(0,1)$.  To that end, denote $H$ to be the Hajek projection of $U$, i.e.,
$$H = \sum_{i=1}^{n}U_i - (n-1)E(U),$$
where $U_i = E(U|\bm{X_i,g_i})$.
It can be seen that $U_i = W_i + c_i,$
where 
\begin{align*}
	W_i &= -\frac{1}{nm_i}\sum_{j=n_1+1,j\ne i}^{n}\alpha_j\sum_{k=1}^{m_i}F_2(X_{ik}) +\frac{\Delta_i}{nm_i}\sum_{j=1,j\ne i}^{n}\frac{1}{m_j}\sum_{k=1}^{m_i}\sum_{h=1}^{m_j}g_{ik}f_{jh}(X_{ik}),
\end{align*}
$\Delta_i= I(i\in \{n_1+1,\ldots,n\})$, 
$f_{jh}(x) = E\left[\frac{1}{2}\left(I(X_{jh}<X_{ik})+I(X_{jh}\le X_{ik})\right)|X_{ik}=x\right]$
and 
\begin{align*}
	c_i &= \sum_{j=1}^{n}\alpha_j - \frac{1}{2n}\left(\sum_{j=1}^{n}\alpha_j(1-\alpha_j) + \bigg(\sum_{j=1}^{n}\alpha_j\bigg)^2 + \sum_{j=1}^{n}\alpha_j\right)\\
	&\hspace{1cm}-\frac{1}{n}\sum_{j=n_1+1,j\ne i}^{n}\sum_{k=1,k\ne i,k\ne j}^{n}\bigg(\alpha_j(1-\alpha_k)p + \frac{\alpha_j\alpha_k}{2}\bigg) - \frac{\alpha_i(n-1)}{n}.
\end{align*}
  We can express
\[
	U-E(U) = r_n + \sum_{i=1}^{n}[W_i-E(W_i)],\]
where $r_n = U-H$.  The proof will be complete if we show that 
\begin{equation}\label{DistnW_i}
\frac{\sum_{i=1}^{n} W_i - E(W_i)}{\sqrt{\sum_{i=1}^{n}Var(W_i)}}\stackrel{d}{\rightarrow}N(0,1)
\end{equation}
and
\begin{equation}\label{Negligib_r_n}
	\frac{r_n}{\sqrt{\sum_{i=1}^{n}Var(W_i)}}\stackrel{p}{\rightarrow}0.
\end{equation}
To show \eqref{DistnW_i}, note that $W_i$'s are bounded, independent and, by Condition \ref{cond:regvar}, \linebreak
$\sum_{i=1}^{n}Var(W_i)\to \infty$.  Therefore, by the Lindeberg's Central Limit Theorem,
$$\frac{\sum_{i=1}^{n} W_i - E(W_i)}{\sqrt{\sum_{i=1}^{n}Var(W_i)}}\stackrel{d}{\rightarrow}N(0,1).$$
To establish \eqref{Negligib_r_n}, since $E(r_n) = 0$,   it suffices to show that 
$$E\bigg(\frac{r_n^2}{\sum_{i=1}^{n}Var(W_i)}\bigg)\rightarrow0.$$  The later will be established if we show that $E(r_n^2)=O(1)$.  It can be seen that $r_n$ can be expressed as
  \[r_n= \frac{1}{n}\sum_{1\le {i_1}<{i_2}\le n}R\left((\bm{X_{i_1},g_{i_1}}),(\bm{X_{i_2},g_{i_2}})\right),\]
where $R$ is a bounded function with zero expectation.  The desired order follows from direct calculation by mutual independence of the clusters  and boundedness of $R$.
\end{proof}


\begin{proof}[Proof of Theorem \ref{thm:ratioconsistency}]
It suffices to show that the quantities $\frac{1}{n}\sum_{i=1}^{n}{Var}(W_i)$ and $\frac{1}{n}\sum_{i=1}^{n} [\widehat{W}_i-\widehat{E}(W_i)]^2$ are asymptotically the equivalent.  This  is eastablished by Propostions  \ref{proof3:prop1} , \ref{proof3:prop2} and \ref{proof3:prop3}.
\end{proof}

\begin{proposition}\label{proof3:prop1}
        \begin{align*}
            \frac{1}{n}\sum_{i=1}^{n} [W_i-E(W_i)]^2 - \frac{1}{n}\sum_{i=1}^{n}Var(W_i) \stackrel{p}{\rightarrow} 0
        \end{align*}
         as $n\rightarrow\infty$.
\end{proposition}
\begin{proof}
Since  $E(W_i-E(W_i))^2\le E(W_i^4) \le 1$ and $\sum_{i=1}^\infty i^{-2}E(W_i-E(W_i))^2 <\infty$, 	the claim follows from Kolmogorov's Strong Law of Large Numbers \citep{SS1993}.
\end{proof}

\begin{proposition}\label{proof3:prop2}
        \begin{align*}
            \frac{1}{n}\sum_{i=1}^{n} [W_i-E(W_i)]^2 - \frac{1}{n}\sum_{i=1}^{n} [W_i-\widehat{E}(W_i)]^2 \stackrel{p}{\rightarrow} 0
        \end{align*}
        as $n\rightarrow\infty$.
    \end{proposition}
\begin{proof}
Note that 
\begin{align*}
	&\frac{1}{n}\sum_{i=1}^{n} [W_i-E(W_i)]^2 - \frac{1}{n}\sum_{i=1}^{n} [W_i-\widehat{E}(W_i)]^2\\
	&=-\frac{1}{n}\sum_{i=1}^{n} [E(W_{i})-\widehat{E}(W_{i})]^2 - \frac{2}{n}\sum_{i=1}^{n} [W_{i}-E(W_{i})][E(W_{i})-\widehat{E}(W_{i})].
\end{align*}
It now suffices to show that both terms on the right hand side converge to $0$ in probability.  After some algebra and simplification, 
\begin{align*}
	\widehat{E(W_l)} - E(W_l)
	&= \frac{p-\widetilde{p}}{n+1}\sum_{i=n_1+1}^{n}\alpha_i
\end{align*}
for any $l$, and thus 
\begin{align*}
	\frac{1}{n}\sum_{i=1}^{n}[\widehat{E(W_{i}) }- E(W_{i})]^2&= \frac{1}{n}\sum_{i=1}^{n}\bigg(\frac{p-\widetilde{p}}{n+1}\sum_{j=n_1+1}^{n}\alpha_j\bigg)^2< (p-\widetilde{p})^2,
\end{align*}
since $\sum_{j=n_1+1}^{n}\alpha_j\le n $.  Therefore, 
$$\frac{1}{n}\sum_{i=1}^{n}[\widehat{E(W_{i})} - E(W_{i})]^2\stackrel{p}{\rightarrow}0,$$
since $(p-\widetilde{p})^2\stackrel{p}{\rightarrow}0$.  On the other hand,
\begin{align*}
	\frac{2}{n}\sum_{l=1}^{n}[W_l-E(W_l)][E(W_l) - \widehat{E(W_l)}]
	&= \{2(p-\widetilde{p})\}\left(\frac{\sum_{i=n_1+1}^{n}\alpha_i}{n+1}\right)\left(\frac{1}{n}\sum_{l=1}^{n}[W_l-E(W_l)]\right),
\end{align*}where each of the factors on the right hand side converges to $0$ in probability. 
\end{proof}
    \begin{proposition}\label{proof3:prop3}
        \begin{align*}
            \frac{1}{n}\sum_{i=1}^{n} [W_i-\widehat{E}(W_i)]^2 - \frac{1}{n}\sum_{i=1}^{n} [\widehat{W}_i-\widehat{E}(W_i)]^2\stackrel{p}{\rightarrow} 0
        \end{align*}
        as $n\rightarrow\infty$.
    \end{proposition}

  \begin{proof} Observe that
    \begin{align}
        &\frac{1}{n}\sum_{i=1}^{n} [W_i-\widehat{E}(W_i)]^2 - \frac{1}{n}\sum_{i=1}^{n} [\widehat{W}_i-\widehat{E}(W_i)]^2 \nonumber \\
        &= -\frac{1}{n}\sum_{i=1}^{n} [\widehat{W}_i-W_i]^2 - \frac{2}{n}\sum_{i=1}^{n} [\widehat{W}_i-W_i][W_i-\widehat{E}(W_i)] \label{eq:diff}.
    \end{align}
   We will show in Propositions \ref{proof3:prop4} and \ref{proof3:prop5} that both terms in \eqref{eq:diff} vanish in probability as $n \to \infty$.  
\end{proof}

    \begin{proposition}\label{proof3:prop4}
        \begin{align*}
            \frac{1}{n}\sum_{i=1}^{n} [\widehat{W}_i-W_i]^2\stackrel{p}{\rightarrow} 0
        \end{align*}
         as $n\rightarrow\infty$.
    \end{proposition}

    \begin{proof}
   For $l=1,\cdots,n_1$,
    \begin{align*}
        (W_l - \widehat{W}_l)^2 &= [m_l(n+1)]^{-2}\bigg(\sum_{i=1}^{n}\alpha_i\bigg)^2\bigg(\sum_{k=1}^{m_l}[F_2(X_{lk}) - G_2(X_{lk})]\bigg)^2\\
        &\leq m_l^{-1}(n+1)^{-2}\bigg(\sum_{i=1}^{n}\alpha_i\bigg)^2\sum_{k=1}^{m_l}[F_2(X_{lk}) - G_2(X_{lk})]^2.
    \end{align*}
  Thus,
    \begin{align}\label{eq:sum_part1}
        \sum_{l=1}^{n_1}(W_l - \widehat{W}_l)^2
        &\leq\bigg(\frac{n_2+n_c}{n+1}\bigg)^2\sum_{l=1}^{n_1}\frac{1}{m_l}\sum_{k=1}^{m_l}[F_2(X_{lk}) - G_2(X_{lk})]^2,
    \end{align}
    since $\sum_{i=1}^{n}\alpha_i\le n_2+n_c$.  For any $l=1,\cdots,n$ and $k=1,\cdots,m_l$, it is clear that $E[\widehat{F}_{2j}(X_{lk})-F_2(X_{lk})]=0$ for any $j=n_1+1,\cdots,n$.  Applying  Fubini's Theorem 
    \begin{align*}
        E[(F_2(X_{lk}) - G_2(X_{lk}))^2] &= \bigg(\frac{1}{n_2+n_c}\bigg)^2\sum_{j=n_1+1}^{n}E[F_2(X_{lk})-\widehat{F}_{2j}(X_{lk})]^2.
    \end{align*}
    Clearly, $[F_2(X_{lk})-\widehat{F}_{2j}(X_{lk})]^2 \le 4$ for any  $j=n_1+1,\cdots,n$, $l=1,\cdots,n$ and $k=1,\cdots,m_l$.  therefore,  
    \begin{align*}
        E[(F_2(X_{lk}) - G_2(X_{lk}))^2] &\le  \frac{4}{n_2+n_c}.
    \end{align*} 
    It then follows that
    \begin{align*}
        E\bigg(\sum_{l=1}^{n_1}(W_l - \widehat{W}_l)^2\bigg) &\le \frac{4n_1(n_2+n_c)}{(n+1)^2}=O(1).
    \end{align*}
    On the other hand, for $l=n_1+1,\cdots,n$,  
    \begin{align*}
        &(W_l - \widehat{W}_l)^2\\
        &= \bigg[\frac{1}{m_l(n+1)}\bigg(\sum_{i=n_1+1,i\ne l}^{n}\alpha_i\sum_{k=1}^{m_l}[F_2(X_{lk}) - G_2(X_{lk})]\\
        &\hspace{1cm} - \sum_{j=1,j\ne l}^{n}\sum_{h=m_l^{(1)}+1}^{m_l}[\alpha_j (F_2(X_{lh})-\widehat{F}_{2j}(X_{lh})) + (1-\alpha_j) (F_1(X_{lh})-\widehat{F}_{1j}(X_{lh}))]\bigg)\bigg]^2\\
        &\le 2m_l^{-1}(n+1)^{-2}\bigg(\sum_{i=1,i\ne l}^{n}\alpha_i\bigg)^2\sum_{k=1}^{m_l}[F_2(X_{lk}) - G_2(X_{lk})]^2 + 4[m_l(n+1)]^{-2}\bigg(\sum_{j=1,j\ne l}^{n}\alpha_j\bigg)^2\cdot\\
        &\hspace{1cm}m_l^{(2)}\sum_{h=m_l^{(1)}+1}^{m_l}\bigg[F_2(X_{lh}) - \sum_{j=1,j\ne l}^{n}\frac{\alpha_j}{\sum_{j=1,j\ne l}^{n}\alpha_j}\widehat{F}_{2j}(X_{lh})\bigg]^2 + 4[m_l(n+1)]^{-2}\\
        &\hspace{1cm}\bigg(\sum_{j=1,j\ne l}^{n}(1-\alpha_j)\bigg)^2m_l^{(1)}\sum_{h=m_l^{(1)}+1}^{m_l}\bigg[F_1(X_{lh}) - \sum_{j=1,j\ne l}^{n}\frac{1-\alpha_j}{\sum_{j=1,j\ne l}^{n}(1-\alpha_j)}\widehat{F}_{1j}(X_{lh})\bigg]^2.
    \end{align*}
    Note that, for any  $h=m_l^{(1)}+1,\cdots,m_l$, 
    \begin{align*}
        &E\bigg[F_2(X_{lh})- \sum_{j=1,j\ne l}^{n}\frac{\alpha_j}{\sum_{j=1,j\ne l}^{n}\alpha_j}\widehat{F}_{2j}(X_{lh})\bigg]^2\\
        &= \sum_{j=1,j\ne l}^{n}E\bigg[\bigg(\frac{\alpha_j}{\sum_{j=1,j\ne l}^{n}\alpha_j}\bigg)^2\bigg([F_2(X_{lh})-\widehat{F}_{2j}(X_{lh})]\bigg)^2\bigg]\\
        &\le \frac{4\sum_{j=1,j\ne l}^{n}\alpha_j^2}{(\sum_{j=1,j\ne l}^{n}\alpha_j)^2},
    \end{align*}
    where the first equality follows from  Fubini's Theorem and independence between clusters.  By symmetry, we also have 
    \begin{align*}
        E\bigg[F_1(X_{lh})- \sum_{j=1,j\ne l}^{n}\frac{1-\alpha_j}{\sum_{j=1,j\ne l}^{n}(1-\alpha_j)}\widehat{F}_{1j}(X_{lh})\bigg]^2 &\le\frac{4\sum_{j=1,j\ne l}^{n}(1-\alpha_j)^2}{[\sum_{j=1,j\ne l}^{n}(1-\alpha_j)]^2}.
    \end{align*}
    After some simplification, it follows that
    \begin{align}
        &E\bigg[\sum_{l=n_1+1}^{n}(W_l-\widehat{W}_l)^2\bigg] \nonumber\\
        &\le 8\bigg(\frac{\sum_{i=1,i\ne l}^{n}\alpha_i}{n+1}\bigg)^2 + \frac{16}{(n+1)^2}\cdot(n_2+n_c)^2+\frac{16}{(n+1)^2}\cdot(n_1+n_c)^2=O(1). \label{eq:sum_part2}
    \end{align}
    Combining \eqref{eq:sum_part1} and \eqref {eq:sum_part2}, 
    \begin{align*}
        E\bigg[\sum_{l=1}^{n}(W_l-\widehat{W}_l)^2\bigg]
        &= E\bigg[\sum_{l=1}^{n_1}(W_l-\widehat{W}_l)^2\bigg] + E\bigg[\sum_{l=n_1+1}^{n}(W_l-\widehat{W}_l)^2\bigg] =O(1).
    \end{align*}
 The remainder part of the proof follows from Markov's Inequality.
\end{proof}
    
 \begin{proposition}\label{proof3:prop5}
  	\begin{align*}
  		\frac{2}{n}\sum_{i=1}^{n} [\widehat{W}_i-W_i][W_i-\widehat{E}(W_i)]\stackrel{p}{\rightarrow} 0
  	\end{align*}
  	as $n\rightarrow\infty$.
  \end{proposition}  

\begin{proof}    
    To prove the claim, note that
    \begin{align*}
        \frac{2}{n}\sum_{i=1}^{n} (W_l-\widehat{W}_l)[W_l-\widehat{E(W_l)}]
        &=\frac{2}{n}\sum_{i=1}^{n} (W_l-\widehat{W}_l)^2 + \frac{2}{n}\sum_{i=1}^{n} (W_l-\widehat{W}_l)[\widehat{W}_l-\widehat{E(W_l)}].
    \end{align*}
    We have already shown in Proposition \ref{proof3:prop4} that  ${n^{-1}}\sum_{i=1}^{n} (W_i-\widehat{W}_i)^2\stackrel{p}{\rightarrow}0$.  Therefore, it  remains to show  $n^{-1}\sum_{i=1}^{n} (W_l-\widehat{W}_l)[\widehat{W}_l-\widehat{E(W_l)}]\stackrel{p}{\rightarrow}0$. For $l=1,\cdots,n_1$, 
    \begin{align*}
        &E|(W_l-\widehat{W}_l)[\widehat{E(W_l)}-\widehat{W}_l]|\\ 
        &\le E\bigg|\bigg(\frac{1}{m_l(n+1)}\sum_{i=n_1+1}^{n}\alpha_i\sum_{k=1}^{m_l}[F_2(X_{lk}) - G_2(X_{lk})]\bigg)\cdot\\
        &\hspace{1cm}\bigg(\frac{1-\widetilde{p}}{n+1}\sum_{i=n_1+1}^{n}\alpha_i - \frac{1}{m_l(n+1)}\sum_{i=n_1+1}^{n}\alpha_i\sum_{k=1}^{m_l}G_2(X_{lk})\bigg)\bigg|\\
        &\le \sqrt{Var\bigg(\frac{1}{m_l(n+1)}\sum_{i=n_1+1}^{n}\alpha_i\sum_{k=1}^{m_l}[F_2(X_{lk}) - G_2(X_{lk})]\bigg)}\cdot\\
        &\hspace{1cm}\sqrt{Var\bigg(\frac{1-\widetilde{p}}{n+1}\sum_{i=n_1+1}^{n}\alpha_i - \frac{1}{m_l(n+1)}\sum_{i=n_1+1}^{n}\alpha_i\sum_{k=1}^{m_l}G_2(X_{lk})\bigg)},
    \end{align*}
   since $E[F_2(X_{lk}) - G_2(X_{lk})] = 0$.  For the first variance term, further simplification gives
   \begin{align*}
       &Var\bigg(\frac{1}{m_l(n+1)}\sum_{i=n_1+1}^{n}\alpha_i\sum_{k=1}^{m_l}[F_2(X_{lk}) - G_2(X_{lk})]\bigg)\\
       &\le \frac{n_2+n_c}{[m_l(n+1)]^2}\sum_{i=n_1+1}^{n}\alpha_i^2m_l\sum_{k=1}^{m_l}Var[F_2(X_{lk}) - G_2(X_{lk})], 
   \end{align*}
   where for any  $k=1,\cdots,m_l$.  It is shown in the proof of Proposition \ref{proof3:prop4} that,
    \begin{align*}
        Var[F_2(X_{lk}) - G_2(X_{lk})] = E[(F_2(X_{lk}) - G_2(X_{lk}))^2]\le \frac{4}{n_2+n_c}.
    \end{align*}
     Hence, it follows that
     \begin{align*}
         &Var\bigg(\frac{1}{m_l(n+1)}\sum_{i=n_1+1}^{n}\alpha_i\sum_{k=1}^{m_l}[F_2(X_{lk}) - G_2(X_{lk})]\bigg)\le \frac{4}{(n+1)^2}\sum_{i=n_1+1}^{n}\alpha_i^2
     \end{align*}
   for the first variance term.  And for the second variance term, 
    \begin{align*}
        &Var\bigg(\frac{1-\widetilde{p}}{n+1}\sum_{i=n_1+1}^{n}\alpha_i - \frac{1}{m_l(n+1)}\sum_{i=n_1+1}^{n}\alpha_i\sum_{k=1}^{m_l}G_2(X_{lk})\bigg)\\
        &\le 2\bigg(\frac{\sum_{i=n_1+1}^{n}\alpha_i}{n+1}\bigg)^2Var(\widetilde{p}) + \frac{2(n_2+n_c)}{[m_l(n+1)]^2}\sum_{i=n_1+1}^{n}\alpha_i^2m_l\sum_{k=1}^{m_l}Var[G_2(X_{lk})],
    \end{align*}
    where for $k=1,\cdots,m_l$, 
    \begin{align*}
         Var[G_2(X_{lk})]&= Var\bigg(\frac{1}{n_2+n_c}\sum_{j=n_1+1}^{n}\widehat{F}_{2j}(X_{lk})\bigg)\\
         &= \bigg(\frac{1}{n_2+n_c}\bigg)^2\sum_{j=n_1+1}^{n}Var[\widehat{F}_{2j}(X_{lk})]\le \frac{1}{n_2+n_c}.
    \end{align*}
    Therefore, 
    \begin{align*}
        &Var\bigg(\frac{1-\widetilde{p}}{n+1}\sum_{i=n_1+1}^{n}\alpha_i - \frac{1}{m_l(n+1)}\sum_{i=n_1+1}^{n}\alpha_i\sum_{k=1}^{m_l}G_2(X_{lk})\bigg)\\
        &\le 2\bigg(\frac{\sum_{i=n_1+1}^{n}\alpha_i}{n+1}\bigg)^2Var(\widetilde{p}) + \frac{2}{(n+1)^2}\sum_{i=n_1+1}^{n}\alpha_i^2.
    \end{align*}
   Consequently,
    \begin{align}
        &\sum_{l=1}^{n_1}E|(W_l-\widehat{W}_l)[\widehat{E(W_l)}-\widehat{W}_l]| \nonumber\\ 
        &\le \sum_{l=1}^{n_1}\sqrt{\bigg(\frac{4}{(n+1)^2}\sum_{i=n_1+1}^{n}\alpha_i^2\bigg)\cdot \bigg[2\bigg(\frac{\sum_{i=n_1+1}^{n}\alpha_i}{n+1}\bigg)^2Var(\widetilde{p}) + \frac{2}{(n+1)^2}\sum_{i=n_1+1}^{n}\alpha_i^2\bigg]}\nonumber\\
        &\le n_1\sqrt{\frac{4(n_2+n_c)}{(n+1)^2}\cdot \bigg[2\bigg(\frac{n_2+n_c}{n+1}\bigg)^2Var(\widetilde{p}) + \frac{2(n_2+n_c)}{(n+1)^2}\bigg]} = O(1).\label{eq:sum_part11}
    \end{align}
  Similarly, for $l=n_1+1,\cdots,n$,
    \begin{align*}
        &E|(W_l-\widehat{W}_l)[\widehat{E(W_l)}-\widehat{W}_l]|\\ 
        &\le Var\bigg[\frac{1}{m_l(n+1)}\bigg(\sum_{i=n_1+1,i\ne l}^{n}\alpha_i\sum_{k=1}^{m_l}[F_2(X_{lk}) - G_2(X_{lk})]\\
        &\hspace{1cm} - \sum_{j=1,j\ne l}^{n}\sum_{h=m_l^{(1)}+1}^{m_l}[\alpha_j (F_2(X_{lh})-\widehat{F}_{2j}(X_{lh})) + (1-\alpha_j) (F_1(X_{lh})-\widehat{F}_{1j}(X_{lh}))]\bigg)\bigg]^{\frac{1}{2}}\cdot\\
        &\hspace{1cm}Var\bigg\{\bigg[\frac{1}{n+1}\bigg[\bigg((1-\alpha_l)(1-\widetilde{p}) + \frac{\alpha_l}{2}\bigg)\sum_{i=n_1+1,i\ne l}^{n}\alpha_i - \alpha_l\sum_{j=1, j\ne l}^{n}\bigg((1-\alpha_{j})\widetilde{p}+\frac{\alpha_{j}}{2}\bigg)\bigg]\\
        &\hspace{1cm} - \frac{1}{m_l(n+1)}\bigg(\sum_{i=n_1+1,i\ne l}^{n}\alpha_i\sum_{k=1}^{m_l}G_2(X_{lk})- \sum_{j=1,j\ne l}^{n}\sum_{h=m_l^{(1)}+1}^{m_l}\widehat{F}_{j}(X_{lh})\bigg)\bigg]\bigg\}^{\frac{1}{2}}.
    \end{align*}
    For the first variance term, 
    \begin{align*}
        &Var\bigg[\frac{1}{m_l(n+1)}\bigg(\sum_{i=n_1+1,i\ne l}^{n}\alpha_i\sum_{k=1}^{m_l}[F_2(X_{lk}) - G_2(X_{lk})]\\
        &\hspace{1cm} - \sum_{j=1,j\ne l}^{n}\sum_{h=m_l^{(1)}+1}^{m_l}[\alpha_j (F_2(X_{lh})-\widehat{F}_{2j}(X_{lh})) + (1-\alpha_j) (F_1(X_{lh})-\widehat{F}_{1j}(X_{lh}))]\bigg)\bigg]\\
        &\le 2\bigg(\frac{1}{m_l(n+1)}\bigg)^2\bigg[8m_l^2\sum_{i=n_1+1,i\ne l}^{n}{\alpha_i^2} + 2(n_2+n_c)\sum_{j=1,j\ne l}^{n}\alpha_j^2m_l^{(2)}\sum_{h=m_l^{(1)}+1}^{m_l}E\bigg([F_2(X_{lh})-\widehat{F}_{2j}(X_{lh})]\bigg)^2\\
        &\hspace{1cm}+2(n_1+n_c)\sum_{j=1,j\ne l}^{n}(1-\alpha_j)^2m_l^{(1)}\sum_{h=m_l^{(1)}+1}^{m_l} E\bigg([F_1(X_{lh})-\widehat{F}_{1j}(X_{lh})]\bigg)^2\bigg]\\
        &\le \frac{16(n_2+n_c)}{(n+1)^2} + \frac{16\alpha_l^2(n_2+n_c)^2}{(n+1)^2} + \frac{16\alpha_l(1-\alpha_l)(n_1+n_c)^2}{(n+1)^2}.
    \end{align*}
The second variance term,
    \begin{align*}
        &Var\bigg\{\bigg[\frac{1}{n+1}\bigg[\bigg((1-\alpha_l)(-\widetilde{p})\bigg)\sum_{i=n_1+1,i\ne l}^{n}\alpha_i - \alpha_l\sum_{j=1, j\ne l}^{n}\bigg((1-\alpha_{j})\widetilde{p}\bigg)\bigg]\\
        &\hspace{1cm}  - \frac{1}{m_l(n+1)}\bigg(\sum_{i=n_1+1,i\ne l}^{n}\alpha_i\sum_{k=1}^{m_l}G_2(X_{lk})- \sum_{j=1,j\ne l}^{n}\sum_{h=m_l^{(1)}+1}^{m_l}\widehat{F}_{j}(X_{lh})\bigg)\bigg]\bigg\}\\
        &\le\frac{2}{(n+1)^2}\bigg[(2\alpha_l-1)\sum_{i=1,i\ne l}^{n}\alpha_i - \alpha_l(n-1)\bigg]^2Var(\widetilde{p}) + \\
         &\hspace{1cm} \frac{4}{[m_l(n+1)]^2}\bigg\{(n_2+n_c)\sum_{j=1,j\ne l}^{n}\alpha_j^2Var\bigg(\sum_{k=1}^{m_l^{(1)}}G_2(X_{lk})\bigg)+\sum_{j=1,j\ne l}^{n}Var\bigg(\sum_{h=m_l^{(1)}+1}^{m_l}\widehat{F}_j(X_{lh})\bigg)\bigg\},
    \end{align*}
where the first term is $O(\frac{1}{n})$ since $Var(\widetilde{p})=O(\frac{1}{n})$, and the second term is at most $8/(n+1)$.   Therefore, 
\begin{align}\label{eq:sum_part12}
        \sum_{l=n_1+1}^{n}E|(W_l-\widehat{W}_l)[\widehat{E(W_l)}-W_l]| = O(\sqrt{n}).
    \end{align}
    Combining \eqref{eq:sum_part11} and \eqref{eq:sum_part12}, 
    \begin{align*}
        E\bigg(\bigg|\sum_{l=1}^{n}(W_l-\widehat{W}_l)[\widehat{E(W_l)}-W_l]\bigg|\bigg)=O(\sqrt{n}).
    \end{align*}
    The remainder of the proof follows by Markov's Inequality. 
    \end{proof}

\section{Simulation Results}\label{app:tables}
In this section we present additional simulation tables that we could not present in the main manuscript due to space limitations.   The settings, the notations  and discussions for these tables are as presented in Section \ref{sec:sim} of the main manuscript.  
\begin{table}[H]
\begin{center}
  \begin{tabular}{||c|c|c|c|c|c|c|c|c|c||} 
 \hline
  \multicolumn{10}{||c||}{$(n_1,n_2,n_c) = (20,10,10)$,  max$(m_i^{(j)})$ = 2}\\
 \hline
  $(\rho_1, \rho_2, \rho_{12})$ & $(\sigma_1^2, \sigma_2^2)$ & $\widetilde{Z}$ & $\widetilde{T}$ & $\widehat{Z}^*$ & $\widehat{Z}$ & $\widetilde{Z}_{H}$ & DS & CKH & $\text{CKH}_{T}$\\[0.2ex]
 \hline
 \multirow{2}{*}{(0.9,0.9,0.1)} & (1,1) & 6.40 & 5.41 & 5.96 & 6.26 & 6.89 & 5.04 & 6.39 & 5.33\\
  & (1,5) & 6.48 & 5.16 & 6.35 & 6.31 & 6.86 & 5.57 & 6.46 & 5.37\\
 \hline
 \multirow{2}{*}{(0.1,0.9,0.9)} & (1,1) & 6.02 & 4.90 & 5.91 & 5.99 & 6.79 & 4.97 & 5.51 & 4.43\\
  & (1,5) & 6.18 & 4.93 & 5.91 & 6.12 & 6.94 & 5.91 & 5.78 & 4.52\\
 \hline
 \multirow{2}{*}{(0.1,0.1,0.9)} & (1,1) & 6.36 & 5.26 & 5.95 & 6.03 & 7.27 & 5.20 & 5.57 & 4.56\\
  & (1,5) & 6.12 & 4.78 & 6.18 & 5.94 & 6.88 & 5.73 & 5.50 & 4.29\\
  \hline
  \multicolumn{10}{||c||}{$(n_1,n_2,n_c) = (20,10,10)$, max$(m_i^{(j)})$ = 9}\\
 \hline
  $(\rho_1, \rho_2, \rho_{12})$ & $(\sigma_1^2, \sigma_2^2)$ & $\widetilde{Z}$ & $\widetilde{T}$ & $\widehat{Z}^*$ & $\widehat{Z}$ & $\widetilde{Z}_{H}$ & DS & CKH & $\text{CKH}_{T}$\\[0.2ex]
 \hline
 \multirow{2}{*}{(0.9,0.9,0.1)} & (1,1) & 6.06 & 5.19 & 6.50 & 5.90 & 6.81 & 4.97 & 5.96 & 4.90\\
  & (1,5) & 6.86 & 5.47 & 6.76 & 6.61 & 7.37 & 5.89 & 6.16 & 5.08\\
 \hline
 \multirow{2}{*}{(0.1,0.9,0.9)} & (1,1) & 6.11 & 4.85 & 6.61 & 6.00 & 7.51 & 5.20 & 5.96 & 4.96\\
  & (1,5) & 6.71 & 5.17 & 6.50 & 6.55 & 7.71 & 6.73 & 6.44 & 4.90\\
 \hline
 \multirow{2}{*}{(0.1,0.1,0.9)} & (1,1) & 6.17 & 5.20 & 6.36 & 5.55 & 8.21* & 5.33 & 6.26 & 5.18\\
  & (1,5) & 6.26 & 5.02 & 6.97 & 5.97 & 8.32* & 5.67 & 6.02 & 4.66\\
  \hline
\end{tabular}\\
{{\caption{\small\label{table4} Empirical type-I error rate (\%) under ignorable cluster size when both groups follow multivariate Gaussian distributions and nominal type-I error rate is set to be 5\%. $n_1$, $n_2$ are the number of incomplete clusters of group 1, and 2, respectively.  $n_c$ is the number of complete clusters.  $\max(m_i^{(j)})$ stands for the maximum allowed intra-cluster group (ICG) sizes.  The parameters $\sigma_1^2$, $\sigma_2^2$, $\rho_1$, $\rho_2$ and $\rho_{12}$ are directly associated with the theoretical scale matrix of the distributions, as mentioned in Section \ref{subsec:simsetup}.  The methods $\widetilde{Z}$, $\widetilde{T}$, $\widehat{Z}^*$, $\widehat{Z}$ and $\widetilde{Z}_{H}$ are the proposed method and its variants.  DS is the test proposed in \cite{DS2005} whereas CKH and ${\rm CKH}_{T}$ are the test by  \citet{CKH2021} and its small-sample approximation.  Design here is unbalanced with slightly more incomplete clusters in group 1.  The superscript $*$ on the numbers in the body of the table indicates at least 1 (out of 10,000 simulations) result is not available due to negative variance estimation.}}}
\end{center}
\end{table}

\newpage
\begin{table}[!htbp]
\begin{center}
  \begin{tabular}{||c|c|c|c|c|c|c|c|c|c||} 
 \hline
  \multicolumn{10}{||c||}{$(n_1,n_2,n_c) = (10,10,20)$, max$(m_i^{(j)})$ = 2}\\
 \hline
  $(\rho_1, \rho_2, \rho_{12})$ & $(\sigma_1^2, \sigma_2^2)$ & $\widetilde{Z}$ & $\widetilde{T}$ & $\widehat{Z}^*$ & $\widehat{Z}$ & $\widetilde{Z}_{H}$ & DS & CKH & $\text{CKH}_{T}$\\[0.2ex]
 \hline
 \multirow{2}{*}{(0.9,0.9,0.1)} & (1,1) & 6.17 & 5.25 & 6.41 & 6.08 & 6.13 & 5.36 & 5.87 & 5.12\\
  & (1,5) & 5.92 & 5.00 & 6.35 & 5.92 & 6.09 & 5.70 & 5.95 & 5.05\\
 \hline
 \multirow{2}{*}{(0.1,0.9,0.9)} & (1,1) & 5.92 & 4.78 & 6.01 & 5.84 & 6.22 & 5.42 & 5.51 & 4.66\\
  & (1,5) & 5.91 & 4.84 & 6.41 & 6.14 & 6.36 & 6.75 & 5.68 & 4.88\\
 \hline
 \multirow{2}{*}{(0.1,0.1,0.9)} & (1,1) & 5.96 & 4.82 & 6.10 & 5.66 & 6.09 & 5.31 & 5.30 & 4.59\\
  & (1,5) & 5.93 & 4.63 & 6.27 & 5.96 & 6.34 & 6.36 & 5.79 & 4.81\\
  \hline
  \multicolumn{10}{||c||}{$(n_1,n_2,n_c) = (10,10,20)$, max$(m_i^{(j)})$ = 9}\\
 \hline
  $(\rho_1, \rho_2, \rho_{12})$ & $(\sigma_1^2, \sigma_2^2)$ & $\widetilde{Z}$ & $\widetilde{T}$ & $\widehat{Z}^*$ & $\widehat{Z}$ & $\widetilde{Z}_{H}$ & DS & CKH & $\text{CKH}_{T}$\\[0.2ex]
 \hline
 \multirow{2}{*}{(0.9,0.9,0.1)} & (1,1) & 6.25 & 5.18 & 6.27 & 5.92 & 5.93 & 4.94 & 5.91 & 5.01\\
  & (1,5) & 6.05 & 5.05 & 6.40 & 5.91 & 6.03 & 5.56 & 5.97 & 5.21\\
 \hline
 \multirow{2}{*}{(0.1,0.9,0.9)} & (1,1) & 5.41 & 4.25 & 5.97 & 5.46 & 5.77 & 4.97 & 5.49 & 4.70\\
  & (1,5) & 5.62 & 4.67 & 6.07 & 5.89 & 6.16 & 6.79 & 5.66 & 4.73\\
 \hline
 \multirow{2}{*}{(0.1,0.1,0.9)} & (1,1) & 5.61 & 4.74 & 6.05 & 4.88 & 5.50 & 4.90 & 5.82 & 5.12\\
  & (1,5) & 5.58 & 4.75 & 6.06 & 5.39* & 5.66 & 5.37 & 5.79 & 4.96\\
  \hline
\end{tabular}\\
{ {\caption{\small\label{table5} Empirical type-I error rate (\%) under ignorable cluster size when both groups follow multivariate Gaussian distributions and nominal type-I error rate is set to be 5\%. $n_1$, $n_2$ are the number of incomplete clusters of group 1, and 2, respectively.  $n_c$ is the number of complete clusters.  $\max(m_i^{(j)})$ stands for the maximum allowed intra-cluster group (ICG) sizes.  The parameters $\sigma_1^2$, $\sigma_2^2$, $\rho_1$, $\rho_2$ and $\rho_{12}$ are directly associated with the theoretical scale matrix of the distributions, as mentioned in Section \ref{subsec:simsetup}.  The methods $\widetilde{Z}$, $\widetilde{T}$, $\widehat{Z}^*$, $\widehat{Z}$ and $\widetilde{Z}_{H}$ are the proposed method and its variants.  DS is the test proposed in \cite{DS2005} whereas CKH and ${\rm CKH}_{T}$ are the test by  \citet{CKH2021} and its small-sample approximation.   Design here is comparatively balanced with more complete clusters.  The superscript $*$ on the numbers in the body of the table indicates at least 1 (out of 10,000 simulations) result is not available due to negative variance estimation.}}}
\end{center}
\end{table}

\newpage
\begin{table}[!htbp]
\begin{center}
\begin{tabular}{||c|c|c|c|c|c|c|c|c|c||} 
  \hline
  \multicolumn{10}{||c||}{$(n_1,n_2,n_c) = (10,10,20)$, max$(m_i^{(j)})$ = 2}\\
 \hline
  $(\rho_1, \rho_2, \rho_{12})$ & $(\sigma_1^2, \sigma_2^2)$ & $\widetilde{Z}$ & $\widetilde{T}$ & $\widehat{Z}^*$ & $\widehat{Z}$ & $\widetilde{Z}_{H}$ & DS & CKH & $\text{CKH}_{T}$\\[0.2ex]
 \hline
 \multirow{6}{*}{(0.1,0.1,0.9)} & (1,1) & 5.77 & 4.60 & 6.06 & 5.52 & 5.88 & 5.35 & 4.92 & 4.12\\
  & (1,5) & 5.79 & 4.57 & 6.28 & 6.09 & 6.30 & 7.24 & 5.53 & 4.75\\
  & (1,25) & 6.11 & 4.57 & 6.40 & 6.25 & 6.51 & 10.67 & 5.83 & 4.76\\
  & (1,50) & 6.14 & 4.69 & 6.51 & 6.34 & 6.65 & 12.10 & 5.75 & 4.79\\
  & (1,75) & 6.26 & 4.72 & 6.55 & 6.51 & 6.74 & 12.51 & 5.77 & 4.75\\
  & (1,100) & 6.28 & 4.81 & 6.52 & 6.58 & 6.69 & 12.77 & 5.74 & 4.70\\
  \hline
    \multicolumn{10}{||c||}{$(n_1,n_2,n_c) = (10,10,20)$, max$(m_i^{(j)})$ = 9}\\
 \hline
  $(\rho_1, \rho_2, \rho_{12})$ & $(\sigma_1^2, \sigma_2^2)$ & $\widetilde{Z}$ & $\widetilde{T}$ & $\widehat{Z}^*$ & $\widehat{Z}$ & $\widetilde{Z}_{H}$ & DS & CKH & $\text{CKH}_{T}$\\[0.2ex]
 \hline
 \multirow{6}{*}{(0.1,0.1,0.9)} & (1,1) & 6.05 & 5.21 & 6.46 & 5.28 & 5.81 & 5.70 & 5.97 & 5.23\\
  & (1,5) & 5.65 & 4.48 & 6.21 & 5.46 & 5.87 & 6.07 & 5.64 & 4.71\\
  & (1,25) & 5.98 & 4.53 & 6.27 & 6.11* & 6.47* & 8.28 & 5.06 & 4.03\\
  & (1,50) & 5.92 & 4.45 & 6.44 & 6.07* & 6.61* & 9.47 & 5.29 & 4.13\\
  & (1,75) & 5.94 & 4.44 & 6.45 & 6.21* & 6.64* & 10.07 & 5.33 & 4.42\\
  & (1,100) & 5.85 & 4.44 & 6.39 & 6.27* & 6.71* & 10.49 & 5.46 & 4.48\\
  \hline
\end{tabular}\\
{ {\caption{\small\label{table6} Empirical type-I error rate (\%) under ignorable cluster size when both groups follow multivariate Gaussian distributions and nominal type-I error rate is set to be 5\%. $n_1$, $n_2$ are the number of incomplete clusters of group 1, and 2, respectively.  $n_c$ is the number of complete clusters.  $\max(m_i^{(j)})$ stands for the maximum allowed intra-cluster group (ICG) sizes.  The parameters $\sigma_1^2$, $\sigma_2^2$, $\rho_1$, $\rho_2$ and $\rho_{12}$ are used to construct the theoretical variance-covariance matrix of the distributions, as mentioned in Section \ref{subsec:simsetup}.  The methods $\widetilde{Z}$, $\widetilde{T}$, $\widehat{Z}^*$,  $\widehat{Z}$ and $\widetilde{Z}_{H}$ are the proposed method and its variants.  DS is the test proposed in \cite{DS2005} whereas CKH and ${\rm CKH}_{T}$ are the test by \cite{CKH2021} and its small-sample approximation.  The increased difference between $\sigma_1^2$ and $\sigma_2^2$ results in gradually more heteroskedasticity, and eventually a noticeable violation of $F_1 = F_2$.  The superscript $*$ on the numbers in the body of the table indicates at least 1 (out of 10,000 simulations) result is not available due to negative variance estimation.}}}
\end{center}
\end{table}

\begin{table}[!htbp]
\begin{center}
\begin{tabular}{||c|c|c|c|c|c|c|c|c|c|c|c|c||} 
 \hline
  \multicolumn{13}{||c||}{$(n_1,n_2,n_c) = (20,10,10)$}\\
  \multicolumn{13}{||c||}{$(\sigma_1^2, \sigma_2^2, \rho_1, \rho_2, \rho_{12}) = (1,1,0.1,0.1,0.9)$}\\
 \hline
  $c_1$ & $c_2$ & $\mu_d$ & $p$ & $p_0$ & $\widetilde{Z}$ & $\widetilde{T}$ & $\widehat{Z}^*$ & $\widehat{Z}$ & $\widetilde{Z}_{H}$ & DS & CKH & $\text{CKH}_{T}$\\[0.2ex]
 \hline
 2 & 2 & 0 & 0.50 & 0.50 & 6.32 & 5.14 & 6.27 & 5.77 & 7.92 & 5.41 & 5.26 & 4.48\\
 \hline
 2 & 3 & -1 & 0.58 & 0.51 & 21.73 & 19.37 & 15.84 & 20.67 & 24.15 & 18.02 & 5.68 & 4.53\\
 \hline
 3 & 2 & 1 & 0.42 & 0.49 & 21.76 & 19.51 & 16.04 & 20.66 & 21.95 & 18.05 & 5.84 & 4.75\\
 \hline
 2 & 4 & -2 & 0.64 & 0.50 & 47.75 & 44.18 & 34.83 & 45.97 & 50.90 & 39.41 & 5.61 & 4.48\\
 \hline
 4 & 2 & 2 & 0.36 & 0.50 & 47.81 & 44.57 & 34.37 & 46.32 & 47.82 & 39.73 & 5.66 & 4.80\\
 \hline
 2 & 5 & -3 & 0.68 & 0.49 & 64.06 & 61.16 & 49.34 & 62.56 & 67.20* & 53.51 & 6.26 & 5.16\\
 \hline
 5 & 2 & 3 & 0.32 & 0.51 & 63.58 & 60.55 & 48.93 & 62.07 & 63.37 & 53.23 & 6.06 & 4.99\\
 \hline
 2 & 6 & -4 & 0.70 & 0.48 & 73.09 & 70.44 & 58.04 & 71.87 & 75.43* & 61.98 & 8.26 & 7.02\\
 \hline
 6 & 2 & 4 & 0.30 & 0.52 & 73.72 & 71.12 & 59.58 & 72.47 & 73.55 & 61.74 & 7.73 & 6.29\\
 \hline
 2 & 7 & -5 & 0.72 & 0.44 & 78.85 & 76.78 & 66.14 & 77.92 & 81.07* & 66.50 & 11.64 & 9.92\\
 \hline
 7 & 2 & 5 & 0.28 & 0.56 & 79.37 & 77.03 & 66.06 & 78.31 & 79.06 & 66.97 & 11.81 & 9.84\\
 \hline
\end{tabular}\\
{ {\caption{\small\label{table9} Achieved power (\%) under informative cluster size for detecting a violation of $p=\frac{1}{2}$ when both groups follow multivariate Cauchy distributions. $n_1$, $n_2$ are the number of incomplete clusters of group 1, and 2, respectively.  $n_c$ is the number of complete clusters.  The parameters $\sigma_1^2$, $\sigma_2^2$, $\rho_1$, $\rho_2$ and $\rho_{12}$ are directly associated with the theoretical scale matrix of the distributions, as mentioned in Section \ref{subsec:simsetup}.  $p$ and $p_0$ are the theoretical values of the WMW treatment effect under informative ($p$) and ignorable cluster size ($p_0$), respectively.  The parameters $c_1$ and $c_2$ control the informativeness of the cluster size.  The case $c_1=c_2$ is noninformative (igorable) cluster size.   $\mu_d$ is the difference in theoretical means between the two groups.  The methods $\widetilde{Z}$, $\widetilde{T}$, $\widehat{Z}^*$, $\widehat{Z}$ and $\widetilde{Z}_{H}$ are the proposed method and its variants.  DS is the test proposed in \cite{DS2005} whereas CKH and ${\rm CKH}_{T}$ are the test by \citet{CKH2021} and its small-sample approximation.  The superscript $*$ on the numbers in the body of the table indicates at least 1 (out of 10,000 simulations) result is not available due to negative variance estimation.}}}
\end{center}
\end{table}

\begin{table}[!htbp]
\begin{center}
\begin{tabular}{||c|c|c|c|c|c|c|c|c|c|c|c|c||} 
 \hline
 \multicolumn{13}{||c||}{$(n_1,n_2,n_c) = (20,10,10)$}\\
  \multicolumn{13}{||c||}{$(\sigma_1^2, \sigma_2^2, \rho_1, \rho_2, \rho_{12}) = (1,1,0.9,0.9,0.1)$}\\
 \hline
  $c_1$ & $c_2$ & $\mu_d$ & $p$ & $p_0$ & $\widetilde{Z}$ & $\widetilde{T}$ & $\widehat{Z}^*$ & $\widehat{Z}$ & $\widetilde{Z}_{H}$ & DS & CKH & $\text{CKH}_{T}$\\[0.2ex]
 \hline
 2 & 2 & 0 & 0.50 & 0.50 & 6.22 & 5.29 & 6.24 & 5.99 & 7.18 & 5.00 & 5.47 & 4.59\\
 \hline
 2 & 3 & -1 & 0.58 & 0.51 & 18.46 & 16.45 & 16.21 & 17.78 & 20.28* & 15.07 & 5.90 & 4.96\\
 \hline
 3 & 2 & 1 & 0.42 & 0.49 & 18.65 & 16.70 & 16.75 & 18.05 & 18.98 & 15.07 & 5.79 & 4.85\\
 \hline
 2 & 4 & -2 & 0.64 & 0.50 & 40.52 & 37.49 & 35.94 & 39.58 & 43.08* & 32.64 & 5.89 & 4.78\\
 \hline
 4 & 2 & 2 & 0.36 & 0.50 & 40.74 & 37.81 & 35.86 & 39.76 & 40.77 & 32.49 & 5.97 & 4.85\\
 \hline
 2 & 5 & -3 & 0.68 & 0.49 & 55.86 & 52.79 & 50.65 & 54.83 & 58.95* & 44.74 & 6.39 & 5.20\\
 \hline
 5 & 2 & 3 & 0.32 & 0.51 & 55.52 & 52.38 & 50.32 & 54.49 & 55.42 & 44.63 & 6.50 & 5.36\\
 \hline
 2 & 6 & -4 & 0.70 & 0.48 & 65.04 & 62.17 & 59.52 & 64.03 & 67.53* & 52.43 & 8.76 & 7.50\\
 \hline
 6 & 2 & 4 & 0.30 & 0.52 & 65.68 & 62.80 & 60.00 & 64.53 & 65.46 & 52.69 & 8.46 & 6.94\\
 \hline
 2 & 7 & -5 & 0.72 & 0.44 & 71.72 & 68.89 & 66.80 & 70.79 & 74.05* & 57.26 & 12.30 & 10.49\\
 \hline
 7 & 2 & 5 & 0.28 & 0.56 & 71.70 & 68.90 & 66.16 & 70.83 & 71.53 & 57.31 & 11.91 & 10.18\\
 \hline
\end{tabular}\\
{ {\caption{\small\label{table10} Achieved power (\%) under informative cluster size for detecting a violation of $p=\frac{1}{2}$ when both groups follow multivariate Cauchy distributions. $n_1$, $n_2$ are the number of incomplete clusters of group 1, and 2, respectively.  $n_c$ is the number of complete clusters.  The parameters $\sigma_1^2$, $\sigma_2^2$, $\rho_1$, $\rho_2$ and $\rho_{12}$ are directly associated with the theoretical scale matrix of the distributions, as mentioned in Section \ref{subsec:simsetup}.  $p$ and $p_0$ are the theoretical values of the WMW treatment effect under informative ($p$) and ignorable cluster size ($p_0$), respectively.  The parameters $c_1$ and $c_2$ control the informativeness of the cluster size.  The case $c_1=c_2$ is noninformative (igorable) cluster size.   $\mu_d$ is the difference in theoretical means between the two groups.  The methods $\widetilde{Z}$, $\widetilde{T}$, $\widehat{Z}^*$, $\widehat{Z}$ and $\widetilde{Z}_{H}$ are the proposed method and its variants.  DS is the test proposed in \cite{DS2005} whereas CKH and ${\rm CKH}_{T}$ are the test by \citet{CKH2021} and its small-sample approximation.  The superscript $*$ on the numbers in the body of the table indicates at least 1 (out of 10,000 simulations) result is not available due to negative variance estimation.  In particular, simulation results from this table uses a slightly different scale matrix compared to Table \ref{table9}.}}}
\end{center}
\end{table}

\newpage
\begin{table}[H]
\begin{center}
\begin{tabular}{||c|c|c|c|c|c|c|c|c|c|c|c||} 
 \hline
  \multicolumn{12}{||c||}{$(n_1,n_2,n_c) = (20,10,10)$}\\
  \multicolumn{12}{||c||}{$(\sigma_1^2, \sigma_2^2, \rho_1, \rho_2, \rho_{12}) = (1,1,0.1,0.1,0.9)$}\\
 \hline
  $c_1$ & $c_2$ & $\mu_d$ & $p$ & $p_0$ & $\widetilde{Z}$ & $\widetilde{T}$ & $\widehat{Z}^*$ & $\widehat{Z}$ & $\widetilde{Z}_{H}$ & CKH & $\text{CKH}_0$\\[0.2ex]
 \hline
 2 & 2 & 0 & 0.50 & 0.50 & 93.68 & 94.86 & 93.73 & 94.23 & 92.08 & 94.74 & 94.74\\
 \hline
 2 & 3 & -1 & 0.58 & 0.51 & 93.93 & 94.81 & 93.77 & 94.35 & 92.55 & 83.97 & 94.37\\
 \hline
 3 & 2 & 1 & 0.42 & 0.49 & 93.54 & 94.50 & 93.38 & 94.08 & 93.25 & 83.18 & 94.05\\
 \hline
 2 & 4 & -2 & 0.64 & 0.50 & 93.57 & 94.48 & 93.04 & 94.01 & 91.62 & 61.88 & 94.47\\
 \hline
 4 & 2 & 2 & 0.36 & 0.50 & 93.64 & 94.88 & 93.51 & 94.30 & 93.73 & 61.66 & 94.29\\
 \hline
 2 & 5 & -3 & 0.68 & 0.49 & 93.90 & 94.92 & 93.65 & 94.29 & 91.56* & 39.42 & 94.26\\
 \hline
 5 & 2 & 3 & 0.32 & 0.51 & 93.46 & 94.36 & 93.14 & 93.87 & 93.42 & 39.10 & 94.43\\
 \hline
 2 & 6 & -4 & 0.70 & 0.48 & 93.47 & 94.32 & 92.92 & 93.89 & 90.45* & 23.19 & 93.38\\
 \hline
 6 & 2 & 4 & 0.30 & 0.52 & 93.37 & 94.38 & 93.26 & 93.86 & 93.52 & 23.08 & 93.90\\
 \hline
 2 & 7 & -5 & 0.72 & 0.44 & 93.68 & 94.60 & 92.81 & 93.96 & 90.40* & 13.68 & 94.27\\
 \hline
 7 & 2 & 5 & 0.28 & 0.56 & 93.01 & 93.92 & 92.39 & 93.35 & 93.11 & 13.51 & 94.64\\
 \hline
\end{tabular}\\
{ {\caption{\small\label{table13} Coverage rate (\%) under informative cluster size for the treatment effect $p$ when both groups follow multivariate Cauchy distributions. $n_1$, $n_2$ are the number of incomplete clusters of group 1, and 2, respectively.  $n_c$ is the number of complete clusters.  The parameters $\sigma_1^2$, $\sigma_2^2$, $\rho_1$, $\rho_2$ and $\rho_{12}$ are directly associated with the scale matrix of the distributions, as mentioned in Section \ref{subsec:simsetup}.  $p$ and $p_0$ are the theoretical values of the WMW treatment effec under informative ($p$) and ignorable cluster size ($p_0$), respectively.  The parameters $c_1$ and $c_2$ control the informativeness of the cluster size.  The case $c_1=c_2$ is noninformative (igorable) cluster size.  $\mu_d$ is the difference in theoretical means between the two groups.  The methods $\widetilde{Z}$, $\widetilde{T}$, $\widehat{Z}^*$, $\widehat{Z}$ and $\widetilde{Z}_{H}$ are the proposed method and its variants.  DS is the test proposed in \cite{DS2005} whereas CKH is the test by \citet{CKH2021}.  ${\rm CKH}_0$ shows the coverage rate of the test by \cite{CKH2021} for covering $p_0$, instead of $p$.  The superscript $*$ on the numbers in the body of the table indicates at least 1 (out of 10,000 simulations) result is not available due to negative variance estimation.}}}
\end{center}
\end{table}

\newpage
\begin{table}[H]
\begin{center}
\begin{tabular}{||c|c|c|c|c|c|c|c|c|c|c|c||} 
\hline
 \multicolumn{12}{||c||}{$(n_1,n_2,n_c) = (20,10,10)$}\\
  \multicolumn{12}{||c||}{$(\sigma_1^2, \sigma_2^2, \rho_1, \rho_2, \rho_{12}) = (1,1,0.9,0.9,0.1)$}\\
 \hline
  $c_1$ & $c_2$ & $\mu_d$ & $p$ & $p_0$ & $\widetilde{Z}$ & $\widetilde{T}$ & $\widehat{Z}^*$ & $\widehat{Z}$ & $\widetilde{Z}_{H}$ & CKH & $\text{CKH}_0$\\[0.2ex]
 \hline
 2 & 2 & 0 & 0.50 & 0.50 & 93.78 & 94.71 & 93.76 & 94.01 & 92.82 & 94.53 & 94.53\\
 \hline
 2 & 3 & -1 & 0.58 & 0.51 & 93.20 & 94.15 & 93.63 & 93.35 & 92.22* & 86.57 & 94.15\\
 \hline
 3 & 2 & 1 & 0.42 & 0.49 & 93.28 & 94.31 & 93.02 & 93.56 & 92.96 & 85.97 & 94.31\\
 \hline
 2 & 4 & -2 & 0.64 & 0.50 & 93.13 & 94.05 & 93.10 & 93.45 & 91.50* & 69.63 & 94.10\\
 \hline
 4 & 2 & 2 & 0.36 & 0.50 & 92.95 & 93.87 & 93.00 & 93.30 & 92.88 & 69.20 & 94.02\\
 \hline
 2 & 5 & -3 & 0.68 & 0.49 & 93.21 & 94.09 & 93.45 & 93.45 & 91.38* & 48.98 & 94.25\\
 \hline
 5 & 2 & 3 & 0.32 & 0.51 & 92.89 & 93.97 & 93.27 & 93.24 & 92.87 & 49.84 & 93.99\\
 \hline
 2 & 6 & -4 & 0.70 & 0.48 & 92.77 & 93.86 & 93.05 & 93.09 & 90.44* & 32.99 & 92.82\\
 \hline
 6 & 2 & 4 & 0.30 & 0.52 & 92.89 & 94.00 & 92.77 & 93.22 & 92.91 & 32.75 & 93.20\\
 \hline
 2 & 7 & -5 & 0.72 & 0.44 & 93.07 & 94.04 & 92.82 & 93.22 & 90.55* & 21.30 & 94.18\\
 \hline
 7 & 2 & 5 & 0.28 & 0.56 & 92.43 & 93.45 & 92.27 & 92.82 & 92.53 & 22.00 & 93.76\\
 \hline
\end{tabular}\\
{ {\caption{\small\label{table14} Coverage rate (\%) under informative cluster size for the treatment effect $p$ when both groups follow multivariate Cauchy distributions. $n_1$, $n_2$ are the number of incomplete clusters of group 1, and 2, respectively.  $n_c$ is the number of complete clusters.  The parameters $\sigma_1^2$, $\sigma_2^2$, $\rho_1$, $\rho_2$ and $\rho_{12}$ are directly associated with the scale matrix of the distributions, as mentioned in Section \ref{subsec:simsetup}.  $p$ and $p_0$ are the theoretical values of the WMW treatment effec under informative ($p$) and ignorable cluster size ($p_0$), respectively.  The parameters $c_1$ and $c_2$ control the informativeness of the cluster size.  The case $c_1=c_2$ is noninformative (igorable) cluster size.  $\mu_d$ is the difference in theoretical means between the two groups.  The methods $\widetilde{Z}$, $\widetilde{T}$, $\widehat{Z}^*$, $\widehat{Z}$ and $\widetilde{Z}_{H}$ are the proposed method and its variants.  DS is the test proposed in \cite{DS2005} whereas CKH is the test by \citet{CKH2021}.  ${\rm CKH}_0$ shows the coverage rate of the test by \cite{CKH2021} for covering $p_0$, instead of $p$.  The superscript $*$ on the numbers in the body of the table indicates at least 1 (out of 10,000 simulations) result is not available due to negative variance estimation.  In particular, simulation results from this table uses a slightly different scale matrix compared to Table \ref{table13}.}}}
\end{center}
\end{table}
